\documentclass[fleqn,usenatbib]{mnras}

\usepackage{newtxtext,newtxmath}

\usepackage[T1]{fontenc}
\usepackage{ae,aecompl}

\usepackage{enumitem}
\usepackage{graphicx}	
\usepackage{amsmath}	
\usepackage{amssymb}	

\title[Global Analysis of the TRAPPIST Ultra-Cool Dwarf Transit Survey]{Global Analysis of the TRAPPIST Ultra-Cool Dwarf Transit Survey}

\author[F. Lienhard et al.]{
F. Lienhard,$^{1}$\thanks{E-mail: fl386@cam.ac.uk}
D. Queloz,$^{1}$
M. Gillon,$^{2}$
A. Burdanov,$^{3,4}$
L. Delrez,$^{1,2,5}$
E. Ducrot,$^{2}$
\newauthor
W. Handley,$^{1}$
E. Jehin,$^{5}$
C. A. Murray,$^{1}$
A. H. M. J. Triaud,$^{6}$
E. Gillen,$^{1}$
A. Mortier,$^{1}$
\newauthor
B. V. Rackham$^{3,7,8}$\\
$^{1}$Cavendish Laboratory, JJ Thomson Avenue, Cambridge CB3 0HE, UK\\
$^{2}$Astrobiology Research Unit, Universit\'{e} de Li\`{e}ge, All\'{e}e du 6 Ao\^{u}t 19, Sart Tilman, 4000, Li\`{e}ge 1, Belgium.\\
$^{3}$Department of Earth, Atmospheric and Planetary Sciences, Massachusetts Institute of Technology, Cambridge, MA 02139, USA\\
$^{4}$Instituto de Astrof\'isica de Canarias, V\'ia L\'actea s/n, 38205 La Laguna, Tenerife, Spain\\
$^{5}$Space Sciences, Technologies and Astrophysics Research (STAR) Institute, Universit\'{e} de Li\`{e}ge, 4000 Li\`{e}ge, Belgium.\\
$^{6}$School of Physics \& Astronomy, University of Birmingham, Edgbaston, Birmingham B152TT, UK\\
$^{7}$Kavli Institute for Astrophysics and Space Research, Massachusetts Institute of Technology, Cambridge, MA 02139, USA\\
$^{8}$51 Pegasi b Fellow\\
}
\date{Accepted 2020 July 9. Received 2020 July 9; in original form 2020 March 23}

\pubyear{2020}

\begin{document}
\label{firstpage}
\pagerange{\pageref{firstpage}--\pageref{lastpage}}
\maketitle

\begin{abstract}\
We conducted a global analysis of the TRAPPIST Ultra-Cool Dwarf Transit Survey -- a prototype of the SPECULOOS transit search conducted with the TRAPPIST-South robotic telescope in Chile from 2011 to 2017 -- to estimate the occurrence rate of close-in planets such as TRAPPIST-1b orbiting ultra-cool dwarfs. 
For this purpose, the photometric data of 40 nearby ultra-cool dwarfs were reanalysed in a self-consistent and fully automated manner starting from the raw images. The pipeline developed specifically for this task generates differential light curves, removes non-planetary photometric features and stellar variability, and searches for transits. It identifies the transits of TRAPPIST-1b and TRAPPIST-1c without any human intervention.
To test the pipeline and the potential output of similar surveys, we injected planetary transits into the light curves on a star-by-star basis and tested whether the pipeline is able to detect them. The achieved photometric precision enables us to identify Earth-sized planets orbiting ultra-cool dwarfs as validated by the injection tests. Our planet-injection simulation further suggests a lower limit of 10 per cent on the occurrence rate of planets similar to TRAPPIST-1b with a radius between 1 and 1.3~$R_\oplus$ and the orbital period between 1.4 and 1.8~days.
\end{abstract}




\section{Introduction}

Ultra-cool dwarfs (UCDs: late M dwarfs and brown dwarfs with effective temperature below 2,700~K, \citealt{Martin1999, Kirkpatrick2005}) make up a significant fraction of all stellar objects in the Galaxy. Counting dwarf stars of spectral type M7 and later including L, T, and Y dwarfs, the census of stars and brown dwarfs by \citealt{kirkpatrick_2012} suggests that UCDs account for about 18 per cent of the stellar and substellar objects within 8~pc of the Sun. UCDs are excellent targets in the search for temperate transiting planets  because of their low mass and size. The habitable zones of UCD stars are 30--100 times closer to their host star than that of the Sun due to their low temperature. Therefore, temperate planets exhibit a comparably short orbital period from one to a few days, which increases the likelihood of observing transits \citep{gillon_trappist_2013}. 
Since the radius of a mature UCD is about ten times smaller than the Sun \citep{Chabrier1997, Dieterich2014}, the transit depth of an Earth-sized planet is of the order of 1 per cent, which is within the detection range of small ground-based telescopes. Additionally, planets orbiting UCDs are optimal targets for the characterisation of their atmospheres' chemical composition via transmission spectroscopy \\ \citep{Kalt2009, dewit_2013}. However, despite their high frequency, the statistics of the planetary population of late M dwarfs and brown dwarfs are poorly understood \citep{delrez_speculoos_2018}. This is mainly due to the low intrinsic brightness of UCDs, which reduces the sample of stars within reach of radial velocity and transit searches. Additionally, UCDs can display a high degree of stellar activity complicating the planet detection via the radial velocity and the transit technique. Knowledge of planet occurrence rates comes mainly from the \textit{Kepler} mission \citep[e.g. ][]{dressing_2015}. However, UCDs were not the primary targets of the \textit{Kepler} mission. Furthermore, some ground-based surveys, such as MEarth \citep{Berta_2012} focusing on M dwarfs, were designed for planets with radii above 2$R_{\oplus}$ at short orbital periods, which turned out to be rare \citep{Berta_2013}.

Radial velocity and transit surveys indicate that tightly packed systems of low-mass planets are common around solar-type stars \citep{howard_2010} and red dwarfs \citep{kirkpatrick_2012,dressing_2013,Hardegree_2019}. UCDs could host a comparable planetary population as simulations by \citet{ormel_2017} suggest. In the present analysis, we test whether this hypothesis is compatible with the results of the TRAPPIST Ultra-Cool Dwarf Transit Survey (TRAPPIST-UCDTS, \citealt{gillon_trappist_2013}), which discovered three temperate Earth-sized planets orbiting the late M dwarf TRAPPIST-1 located 12.1 pc from the Sun \citep{gillon_trappist_2016}. Further transit observations and analysis of transit timing variations have revealed that the system hosts 7 rocky, Earth-sized planets in almost perfectly edge-on and circular orbits \citep{gillon_trappist_2017,luger_2017}. No additional planets have been found around TRAPPIST-1 \citep{Ducrot_2020} or any of the other stars in this data set.

The data analysed in this paper was acquired with TRAPPIST-South (TRAnsiting Planets and PlanetesImals Small Telescope-South) \citep{gillon_trappist_2011, jehin_2011}, which is a robotic f/8 Ritchey-Chr\'{e}tien 60 cm telescope located at the La Silla Observatory in Chile. TRAPPIST-UCDTS has been monitoring late M dwarfs and brown dwarfs since 2011 as a prototype survey for the more ambitious SPECULOOS survey \citep{burdanov_2018,delrez_speculoos_2018, gillon2018}.
Here, we reanalyse the data of 40 targets observed between 2011 and 2017 by TRAPPIST-South starting from the raw images. The pipeline created for this analysis was developed alongside the SPECULOOS Southern Observatory (SSO) pipeline \citep{Murray_2020}.

The paper is structured as follows. In Section \ref{sampledescription}, we describe the sample used for this analysis. In Section \ref{Methods}, we describe the light curve detrending, removal of data points affected by non-planetary signals and the modelling of stellar variability and activity. In Section \ref{pipeline}, the light curve of TRAPPIST-1, and thus the performance of the pipeline for the only planetary system found in the data set, is discussed. In Section \ref{flareenergies}, we estimate the energy of the flare events that we find in the light curves. The transit injection tests characterising the statistics of the survey are outlined in Section \ref{injectiontests}, while the results of the injection tests are described in Sections \ref{expnr} and \ref{expnr2}. Finally, we summarise our results, compare them to other surveys, and discuss the significance for multi-planetary systems around UCDs in Section \ref{conclusion}.

\section{TRAPPIST-UCDTS: instrumentation and methodology} \label{sampledescription}

\begin{figure}
    \centering
    \includegraphics[width=\columnwidth]{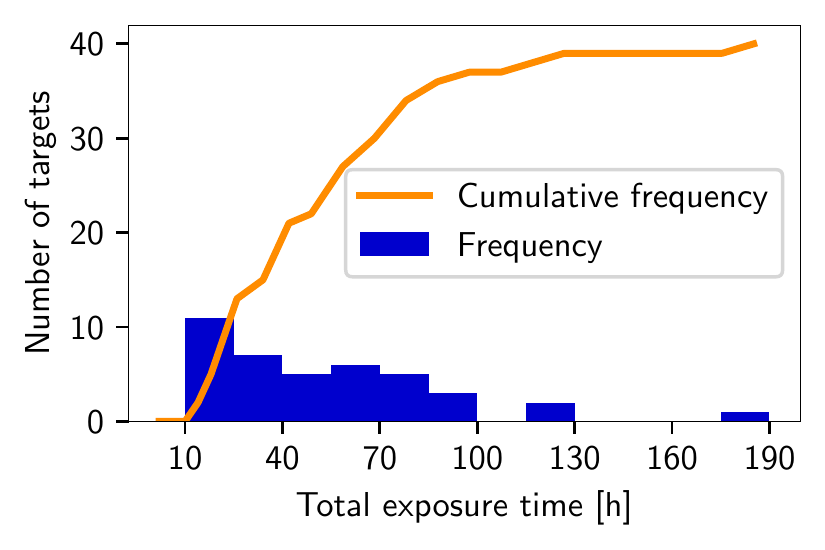}
    \caption{Histogram of the added exposure time of all targets. The target with 175 hours total exposure time is TRAPPIST-1, which was observed more extensively due to the detection of a planetary system around it.}
    \label{fig:obsfreq}
\end{figure}

\begin{figure}
    \centering
    \includegraphics[width=\columnwidth]{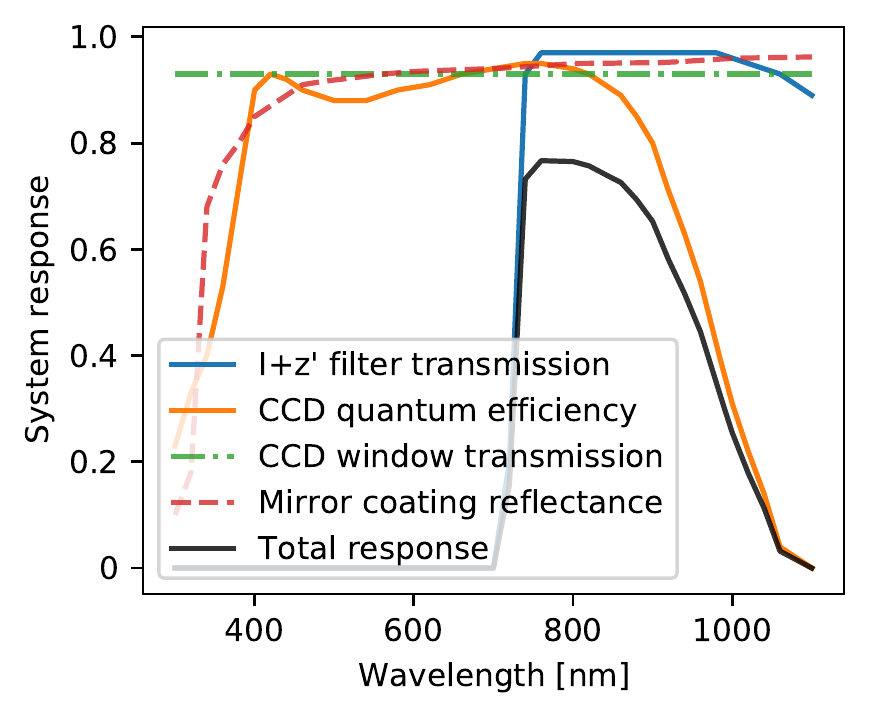}
    \caption{Response function of the TRAPPIST instrument. The total response is computed by multiplying the different components (the mirror coating reflectance has to be taken into account twice). The limiting contributor on the lower end is the filter transmission while the upper end is determined by the mean CCD quantum efficiency.}
    \label{fig:response}
\end{figure}

TRAPPIST-UCDTS was conducted mostly in the I+z$'$ filter due to the redness of the targets (cf. figure 6 in \citealt{delrez_speculoos_2018}). The response function of the TRAPPIST instrument is depicted in Fig. \ref{fig:response} (M. Gillon -- priv. comm.). The field of view is equal to 22\arcmin~$\times$~22\arcmin with a resolution of 0.65 arcsec/pixel. More information can be found in \cite{gillon_trappist_2011}.\footnote{More information about the equipment is also available here: \url{https://www.trappist.uliege.be/cms/c_5288613/en/trappist-equipment}.}
As TRAPPIST-UCDTS is a targeted survey, the exposure time is optimised for each target to maximise the signal-to-noise ratio and duty cycle and is typically between 30 and 60 seconds. Most parts of the light curves allow the detection of transits caused by Earth-sized planets. The UCDs have typically been observed for 50 hours, as visible in Fig. \ref{fig:obsfreq}. The here quoted value is the total exposure time, not the total observation time dedicated to this target. A description of the survey and the first 20 light curves (5 M6, 6 M7, 4 M8, and 5 M9) is provided in \cite{gillon_trappist_2013}. The first 40 light curves observed between 2011 and 2017 are characterised further in \cite{delrez_speculoos_2018} and \citet{burdanov_2018}.
The stars, and associated parameters based on Section \ref{radiiandmasses}, are listed in Table \ref{startable}.
Some of the target stars have been further observed within the TRAPPIST-UCDTS or SPECULOOS programs without identifying additional transiting planets.

\begin{table*}
\caption{The coordinates are in J2015.5. The apparent $K_S$ magnitudes were retrieved from 2MASS \citep{cutri2003} and the $G-G_{RP}$ colour index from Gaia DR2 \citep{gaia1,gaia2}. Radii and masses are estimated as described in Section \ref{radiiandmasses}. The dagger symbol indicates that the absolute $K_S$ magnitude was above 10, and hence we set the mass estimate to the approximate mass of an M9 star. The number of observation nights is shown below \textit{Nights}, while \textit{DP} indicates the number of measurements before bad-weather removal. \textit{Exp} is the mean exposure time and \textit{Span} the number of days between the first and the last observation. In the column below \textit{RMS}, we list the median nightly RMS of the differential light curves before division by the GP mean (i.e. as in Fig. \ref{star1} panel 5).}
\label{startable}
\begin{tabular}{lrrrrrlrrrrr}
\hline
Star id                 & RA & DEC & $K_S$ & $G-G_{RP}$ & Radius  & Mass & Nights & DP & Exp  & Span & RMS  \\
 & {[}deg{]} & {[}deg{]} &  & & {[}$R_\odot${]} & {[}$M_\odot${]}& & & {[}s{]} & {[}d{]} & {[}\%{]}\\
\hline
2MASS J03111547+0106307 & 47.8150  & 1.1085   & 9.7550  & 1.3880 & 0.166 & 0.130            & 6  & 3056  & 28.8 & 387  & 0.49 \\
2MASS J03341218-4953322 & 53.5667  & -49.8902 & 10.3920 & 1.6013 & 0.108 & 0.075 $\dagger$ & 21 & 5953  & 51.8 & 1448 & 0.49 \\
2MASS J07235966-8015179 & 110.9878 & -80.2518 & 10.4400 & 1.4577 & 0.128 & 0.104           & 18 & 4294  & 51.2 & 1151 & 0.38 \\
2MASS J08023786-2002254 & 120.6600 & -20.0428 & 9.5750  & 1.2125 & 0.287 & 0.273           & 5  & 4450  & 20.3 & 1400 & 0.31 \\
2MASS J11592743-5247188 & 179.8564 & -52.7891 & 10.3220 & 1.6090 & 0.108 & 0.075 $\dagger$ & 29 & 8878  & 50   & 38   & 0.71 \\
2MASS J15072779-2000431 & 226.8663 & -20.0123 & 10.6610 & 1.5500 & 0.157 & 0.124           & 5  & 1172  & 50   & 9    & 0.54 \\
2MASS J15345704-1418486 & 233.7331 & -14.3151 & 10.3050 & 1.6171 & 0.110 & 0.075 $\dagger$ & 14 & 4994  & 65   & 760  & 0.39 \\
2MASS J20392378-2926335 & 309.8508 & -29.4461 & 10.3670 & 1.4521 & 0.140 & 0.112           & 12 & 5368  & 55   & 756  & 0.35 \\
2MASS J21342228-4316102 & 323.5938 & -43.2730 & 9.6850  & 1.3983 & 0.170 & 0.134           & 4  & 2495  & 25.9 & 449  & 0.65 \\
2MASS J22135048-6342100 & 333.4614 & -63.7020 & 9.9380  & 1.4442 & 0.134 & 0.108           & 11 & 5533  & 41.2 & 738  & 0.74 \\
APMPM J2330-4737        & 352.5638 & -47.6167 & 10.2790 & 1.5189 & 0.120 & 0.096           & 9  & 3749  & 40.1 & 741  & 0.46 \\
APMPM J2331-2750        & 352.8410 & -27.8272 & 10.6510 & 1.5543 & 0.112 & 0.079           & 4  & 1727  & 24.4 & 1543 & 0.74 \\
DENIS J051737.7-334903  & 79.4094  & -33.8190 & 10.8320 & 1.5912 & 0.118 & 0.093           & 20 & 5416  & 50   & 34   & 0.65 \\
DENIS J1048.0-3956      & 162.0541 & -39.9395 & 8.4470  & 1.5876 & 0.108 & 0.075 $\dagger$ & 8  & 7505  & 18.7 & 734  & 0.49 \\
GJ   283 B              & 115.0859 & -17.4151 & 9.2910  & 1.4724 & 0.124 & 0.101           & 9  & 7630  & 17.7 & 745  & 0.35 \\
GJ 644 C                & 253.8934 & -8.3984  & 8.8160  & 1.5169 & 0.117 & 0.091           & 13 & 12513 & 17.2 & 1130 & 0.41 \\
LEHPM 2- 783            & 304.9574 & -58.2801 & 9.7150  & 1.4957 & 0.164 & 0.129           & 17 & 13311 & 20.7 & 745  & 0.51 \\
LHS  1979               & 121.4061 & -9.5465  & 9.4430  & 1.3129 & 0.182 & 0.144           & 8  & 7973  & 10.8 & 1822 & 0.47 \\
LHS  5303               & 238.1869 & -26.3891 & 9.3150  & 1.4505 & 0.136 & 0.109           & 9  & 6866  & 24.6 & 649  & 0.35 \\
LP  593-68              & 57.7502  & -0.8812  & 10.2320 & 1.5153 & 0.127 & 0.103           & 8  & 2933  & 50.7 & 1096 & 0.45 \\
LP  655-48              & 70.0984  & -5.5017  & 9.5450  & 1.5314 & 0.121 & 0.097           & 4  & 1726  & 45   & 376  & 0.67 \\
LP  666-9               & 133.3984 & -3.4931  & 9.9420  & 1.5754 & 0.109 & 0.075 $\dagger$ & 19 & 6698  & 50.6 & 1465 & 0.63 \\
LP  698-2               & 323.1245 & -5.2012  & 10.3790 & 1.4241 & 0.151 & 0.120            & 5  & 1365  & 55   & 427  & 0.31 \\
LP  760-3               & 337.2250 & -13.4266 & 9.8430  & 1.4771 & 0.119 & 0.094           & 7  & 3418  & 17.7 & 800  & 0.47 \\
LP  775-31              & 68.8180  & -16.1145 & 9.3520  & 1.5455 & 0.134 & 0.108           & 6  & 3065  & 29.2 & 372  & 0.42 \\
LP  787-32              & 140.6950 & -15.7911 & 10.0510 & 1.3439 & 0.166 & 0.130            & 12 & 7877  & 32.4 & 1464 & 0.59 \\
LP  789-23              & 151.6317 & -16.8898 & 10.9920 & 1.5166 & 0.124 & 0.101           & 13 & 1713  & 50   & 18   & 0.60 \\
LP  851-346             & 178.9268 & -22.4171 & 9.8810  & 1.5426 & 0.118 & 0.093           & 10 & 8652  & 27   & 1077 & 0.54 \\
LP  911-56              & 206.6901 & -31.8231 & 10.0380 & 1.4809 & 0.129 & 0.104           & 11 & 5938  & 48.6 & 753  & 0.33 \\
LP  914-54              & 224.1570 & -28.1671 & 8.9280  & 1.5337 & 0.118 & 0.093           & 18 & 15823 & 15.4 & 1108 & 0.48 \\
LP  938-71              & 15.7207  & -37.6277 & 10.0690 & 1.5672 & 0.116 & 0.089           & 7  & 2031  & 43.6 & 488  & 0.45 \\
LP  944-20              & 54.8985  & -35.4276 & 9.5480  & 1.5922 & 0.108 & 0.075 $\dagger$ & 10 & 3925  & 50.1 & 349  & 0.47 \\
LP  993-98              & 40.5279  & -41.4100 & 10.5500 & 1.3524 & 0.168 & 0.132           & 6  & 2935  & 40   & 7    & 0.36 \\
LP 888-18               & 52.8763  & -30.7125 & 10.2640 & 1.5688 & 0.116 & 0.089           & 8  & 3821  & 38.8 & 739  & 0.43 \\
SCR J1546-5534          & 236.6718 & -55.5809 & 9.1120  & 1.5682 & 0.124 & 0.100             & 33 & 14557 & 30   & 103  & 0.35 \\
SIPS J1309-2330         & 197.3411 & -23.5116 & 10.6690 & 1.5578 & 0.116 & 0.089           & 13 & 4706  & 50   & 446  & 0.62 \\
TRAPPIST-1              & 346.6264 & -5.0435  & 10.2960 & 1.5484 & 0.115 & 0.087           & 64 & 12357 & 55   & 935  & 0.48 \\
UCAC4 379-100760        & 271.4345 & -14.3784 & 8.8610  & 1.3889 & 0.145 & 0.116           & 7  & 7067  & 15   & 405  & 0.49 \\
V* DY Psc               & 6.1023   & -1.9716  & 10.5390 & 1.6024 & 0.111 & 0.075 $\dagger$ & 5  & 1136  & 50   & 10   & 0.78 \\
VB 10                   & 289.2277 & 5.1632   & 8.7650  & 1.1037 & 0.113 & 0.083           & 8  & 9505  & 15.3 & 381  & 0.46 \\
\hline
\end{tabular}
\end{table*}


\section{Methods} \label{Methods}
During the survey, the data were inspected for transits by eye on a day-by-day basis. We perform an automated global analysis for this study, which ensures consistency in how the data are treated and which enables us to estimate the probability of finding planets around UCDs using a clearly defined detection criterion.
The individual steps in the automated global analysis are detailed in the following subsections.

\subsection{Image reduction}
To calibrate the science images, we employ the standard offset, bias, and dark subtraction followed by flat-field division.
Thereafter, we perform aperture photometry using the CASUTools software \citep{casutools}. For each target, 20 suitable images are stacked with CASUTools $imstack$ to generate a high-quality image of the field and extract the coordinates ($\text{RA}_i$, $\text{DEC}_i$) of the stars using CASUTools $imcore$, CASUTools $wcsfit$ and $astrometry.net$ \citep{astrometrynet}.

The flux of the stars in all images of the respective field is then computed with $imcorelist$ by adding the pixel values within apertures at the coordinates $\text{RA}_i$ and $\text{DEC}_i$, and subtracting the background flux. To estimate the latter, $imcore$ divides the image into a grid of scale size 64 pixels and computes the interatively k-sigma clipped median on the grid sections. It then lightly filters these values and computes the background via bilinear interpolation. We account for proper motion by recomputing the coordinates of the stars in a target field for each night based on Gaia DR2 \citep{gaia1,gaia2} proper motion data. This is necessary since the target stars are nearby and the time span between the first and the last measurement can be extensive, as visible in Table \ref{startable} or Fig. \ref{star2}. The aperture radius is set to 7.07 pixels (4.6 arcsec) for all stars and images. \cite{Damasso_2010} suggest an aperture radius between 2 and 3 times the mean FWHM of the field stars, which indicates that an aperture radius between 6 and 9 pixels is optimal for most of our science images. To further test this, we computed the RMS of the differential target light curves for different aperture radii. We did not remove bad-weather data, but we applied an iterative global 4-sigma clipping to remove outliers. Based on this analysis, we concluded that 7.07 pixels is the most suitable aperture radius yielding the lowest mean RMS. We chose a fixed aperture over a dynamic one to avoid adding structure to the light curves, which could result from a suboptimal background estimate that a dynamic aperture includes to a varying extent.

\subsection{Differential photometry}
To correct for photometric effects of non-astrophysical origin, we divide the raw target light curve by the weighted average (hereafter called the Artificial Light Curve, ALC) of the median-normalised light curves of selected comparison stars in the same frame as the target star.

First, we define a set of potential reference stars in the same field as the target star. This set consists of all the stars with more than 100 captured electrons per second in their aperture. We also remove bright stars which lead to saturated pixels. We then arrange the median-normalised light curves in a matrix \textit{M} of dimension ($n$,$m$) with $n$ the number of potential reference stars and $m$ the number of science images of the respective field. The matrix form enables us to compare the flux measurements in the same science image (the first axis) or in time (the second axis).

\begin{enumerate}[leftmargin=.5cm]

\item 
The initial step consists of 3-sigma clipping along the first axis of \textit{M}. 
The fluxes of the normalised light curves in the same science image are similarly affected by the airmass, and they are median-normalised along the second axis in \textit{M}. Therefore, we can compare them to each other to remove outliers and to find stars that behave very differently compared to the median star in the field. The standard deviation of the normalised flux values in a science image is dominated by the fainter stars. Consequently, this step represents a coarse clipping on the brighter stars. These bright stars get a high weight in step (iii), which motivates us to sigma clip here since stars can display a high degree of variability which cannot be removed by sigma clipping along the second axis in matrix \textit{M}.

\item 
A reference star's light curve is discarded if more than 20 per cent of its values are removed in step (i).

\item 
We set the statistical weight $w^j$ of each remaining potential reference star $j$ to its median brightness before normalisation:
\begin{equation} \label{eq:weightinitial}
w^j_{\text{initial}}=\text{Md}_i(N^j_i),
\end{equation}
where $N^j_i$ is the photon count for star $j$ in science image $i$ in a fixed aperture and $\text{Md}_i(N^j_i)$ is the median over all photon counts of star $j$.

By setting the statistical weights $w^j$ to the initial weights $w^j_{\text{initial}}$ as above, we compute the ALC as the weighted mean of the normalised light curves:
\begin{equation} \label{eq:alc}
\text{ALC}_i=\sum\limits_{j=1}^{n_{\text{ref}}} \frac{N^j_i w^j}{\text{Md}_i(N^j_i)}  {\left(\sum\limits_{j=1}^{n_{\text{ref}}} w^j\right)}^{-1},
\end{equation}
where $\text{ALC}_i$ is the ALC value for the $i^{\text{th}}$ science image.
We tested a more sophisticated noise model including read-out, dark, and scintillation noise as initial weights. This approach led to the same final weights.
\item 
By dividing each light curve in \textit{M} by the ALC, we get a first estimate of the reference stars' differential light curves. We set the weight of each reference star to the inverse square of the RMS of its differential light curve:
\begin{equation} \label{eq:weights}
w^j={\text{RMS}_i\left(  \frac{N^j_i}{\text{Md}_i(N^j_i)} \frac{1}{\text{ALC}_i}  \right)}^{-2}
\end{equation}
and recompute the ALC as in Eq. \ref{eq:alc}.
\item
We repeat step (iv) until the weights converge to within 0.00001, which is typically achieved after 20 iterations.
Reusing the variability estimate to refine the weights in the ALC calculation is conceptually similar to the detrending approach in \citet{broeg_2005}.

\item 
Some stars can have a higher RMS than similarly bright stars either due to continuous brightness variations caused by, for example, pulsation or binarity or because some parts of the light curve are noisy as a result of bad pixels, an imperfect flat-field correction, or flares. Other stars are perfectly quiet throughout the light curve. In the case of non-stationary noise which is not present in the other light curves, we overweight the noisy parts of the respective light curve and underweight the parts with low temporal RMS by assigning fixed weights in the ALC calculation (cf. Eq. \ref{eq:alc}). Thus, these variable stars can induce some spurious features in the ALC, but they can easily be identified and removed. For this purpose, we modelled the RMS of the normalised light curve of a calm star with median photon count $N^j$ by fitting the following function with three fitting parameters ($a$, $b$, $c$) to the RMS of the stars with the lowest RMS in a given flux bin:

\begin{eqnarray}
RMS(N^j) \ &\propto & \  \frac{\sqrt{\sigma_{\text{photon}}^{2} + \sigma_{\text{const}}^{2}  + \sigma_{\text{scintillation}}^{2}}}{N^j}\nonumber \\ &\propto & \  a \sqrt{\frac{1}{N^j} + \frac{b}{(N^j)^{2}} + c}.
\label{lowerrorfit}
\end{eqnarray}

Poisson noise is proportional to $\sqrt{N^j}$, while scintillation noise is proportional to ${N^j}$ \citep{young1967}. In addition, there is a constant error term to account for read-out and dark current noise.

Stars with an RMS 20 per cent above the minimal RMS of a calm star, calculated in Eq. (\ref{lowerrorfit}), are identified as variable stars. We exclude these stars from the reference star set, as shown in Fig. \ref{excludenoisystars}. The theoretical noise estimate, computed as in \cite{Damasso_2010} eq. 3, is included in Fig. \ref{excludenoisystars} to compare to our fit of the RMS of the entire light curves including unfavourable nights.

\begin{figure}
	\includegraphics[width=\columnwidth]{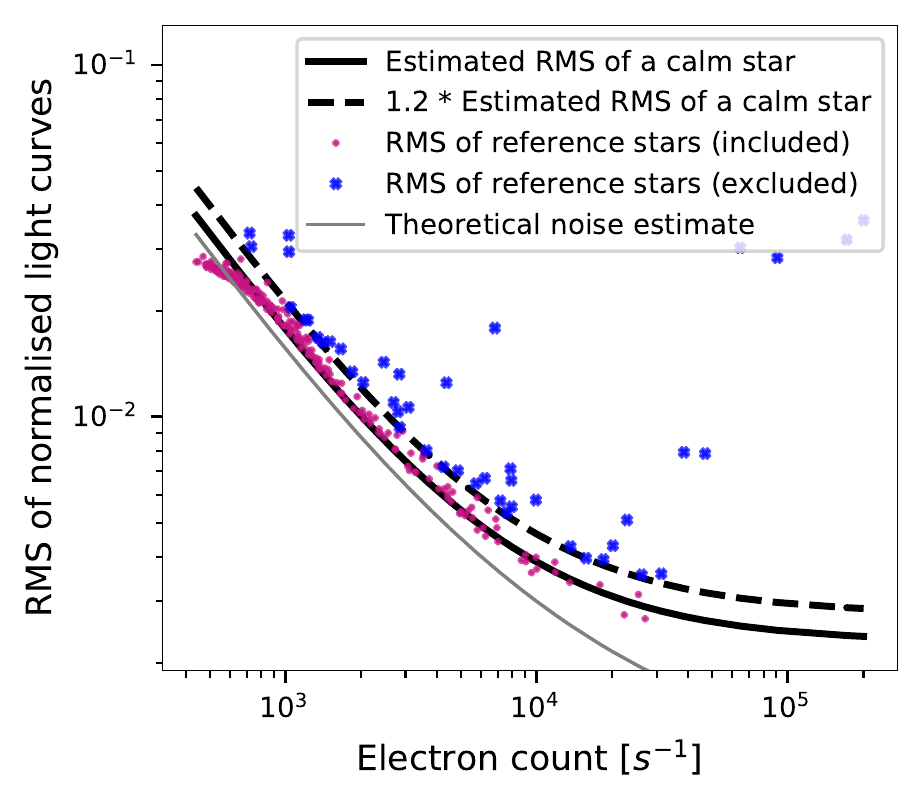}
    \caption{Each dot displays the RMS of a potential reference star's entire light curve in the field of one target star. We used equation \ref{lowerrorfit} as a fitting function to estimate a lower limit for the RMS achievable in this survey as a function of the photon count per second (thick solid black line). If the RMS of the normalised differential light curve of a reference star exceeds the dashed line, we assume that this star is variable and we exclude it from the Artificial Light Curve calculation. Stars with count rates of less than 100 electrons per second have already been excluded for this figure. The RMS does not follow the RMS model for stars with low electron counts since we sigma-clip the flux values measured in the same science image. The RMS values are above the theoretical noise estimate (thin solid black line - \citet{Damasso_2010}) because we compute the RMS of the entire light curves including measurements impacted by unfavourable weather.}
    \label{excludenoisystars}
\end{figure}

\item 
Using the estimate of the ALC from step (v), we can 4-sigma clip along the reference stars' differential light curves to remove any outliers which have not been flagged in step (i). If more than 20 per cent of a reference star's light curve is flagged, its weight is set to zero.
\item 
We repeat steps (v) and (vi).
\item 

The previous steps optimise the ALC with respect to a typical star in the given target field, neglecting the fact that noise can be locally correlated and that light curves are colour-dependent. To improve the correction of the target star, we increase the statistical weight of reference stars near the target star as a function of angular distance to the target star. More specifically, we multiply the statistical weight of the reference stars by a factor $w_\text{d}$ which depends on the angular distance of the respective star to the target star. The distance weighting is computed for each target star field individually, and it remains the same over the full light curve. Since the distance weighting function should be flat near the target star and decay further out, we use the following functional form for $w_\text{d}$:

\begin{equation}
w_\text{d} = \frac{1}{1+\left(a \frac{d_r}{\text{max}_r(d_r)}\right)^2}
\label{distanceweighting}
\end{equation}
where $a$ is a fitting parameter, $d_r$ is the angular distance between target and reference star, and $\text{max}_r(d_r)$ is the angular distance to the star furthest from the target star.

Simultaneously with parameter $a$ in equation (\ref{distanceweighting}), we set a lower threshold for the colour since the target stars are very red compared to the average field star. This cut-off reduces the impact of colour-dependent atmospheric effects on the differential light curves. For this purpose, we calculate the RMS of the final light curves for a grid of potential values of parameter $a$ and colour thresholds based on the Gaia DR2 $G-G_{RP}$ colour index. We proceed with the combination of the two parameters which minimises the RMS of the respective light curve.
 The typical cutoff value for $G-G_{RP}$ is around 0.5, while the parameter $a$ in the distance weighting scheme is more variable but typically around 2.4.

\end{enumerate}

\subsection{Removal of bad-weather data} 
\begin{figure}
	\includegraphics[width=\columnwidth]{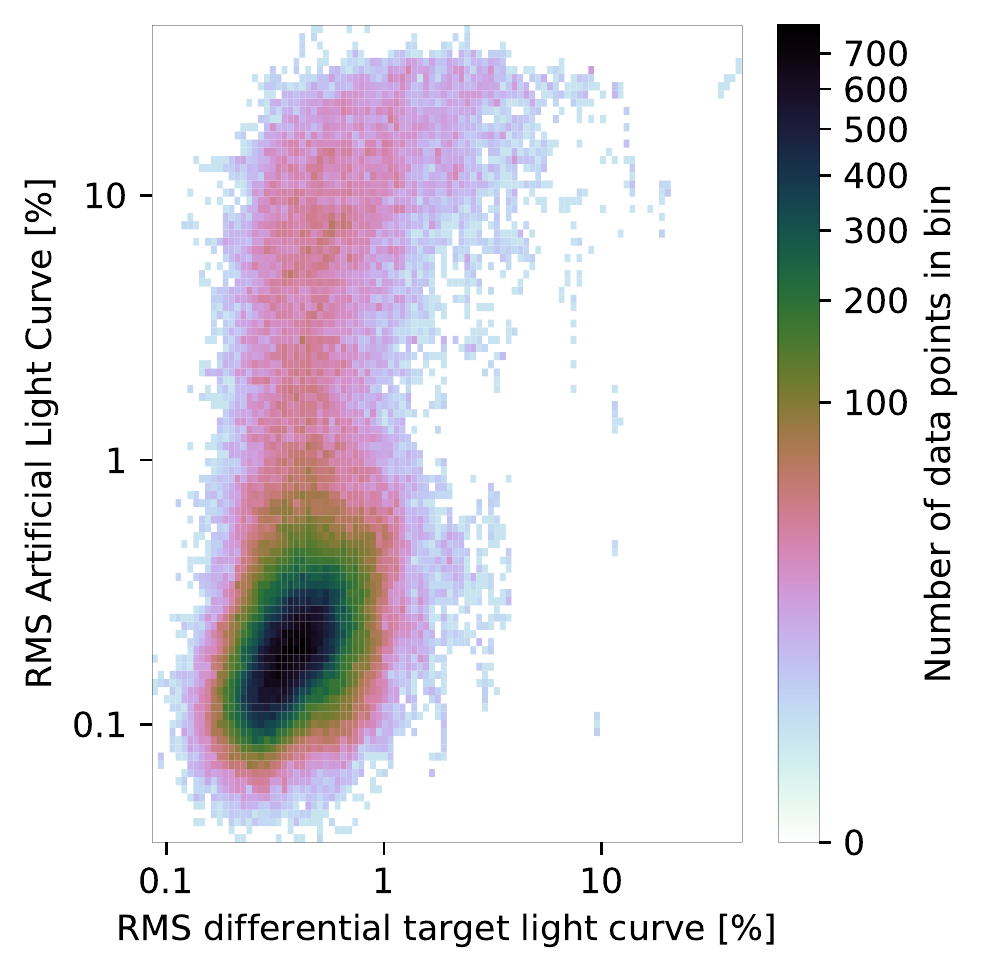}
    \caption{Comparison between the running RMS of the differential target light curve (not GP-detrended but corrected for flares and cosmic rays) and the ALC. The RMS of the target light curve is low if the conditions are favourable. Adverse conditions such as clouds or hazes, identified by the high RMS in the Artificial Light Curve (ALC), can cause some scatter in the target light curve. We remove any flux value where the RMS of the ALC is above 10 per cent. \textsc{cubehelix} colour scheme \citep{Green_2011}.}
    \label{rmsrms}
\end{figure}

Clouds or hazes can lead to spatial inhomogeneity in the transparency of the atmosphere, which hampers the effectiveness of differential photometry. Additionally, these and similar effects decrease the number of photons that reach the detector, thus increasing photon noise. We tackle this problem by removing all flux measurements where the running mean of the ALC is below 0.5 or where the running RMS exceeds 10 per cent. For the running RMS, we compute the RMS of typically 15 flux values closest to the point in time where we intend to estimate the scatter in the light curve. The same procedure is applied for the running mean.\\ Fig. \ref{rmsrms} exemplifies that the running RMS of the differential target light curve is increased if the running RMS of the ALC is high. It also shows that the minimal RMS of the target light curve increases as a function of the ALC RMS, which is consistent with the findings of \cite{Berta_2012}.

\subsection{Removal of flares} 
Flares are characterised by an abrupt rise followed by a slower decrease in brightness.
They can pose significant problems for all of the following light curve optimisation steps, and they tend to confound the transit search. Therefore, it is beneficial to find and remove data points affected by flares. Various approaches were chosen in other surveys to address this problem. \cite{Berta_2012}, for instance, performed a grid search on their data modelling a flare as a fast rise in flux followed by an exponential decay. They removed a night's light curve if they found a flare with an amplitude above 4 sigma. Since stars were often observed during one full night in TRAPPIST-UCDTS, we cannot discard a whole night if there was a flare because it would mean losing a significant amount of data. The MEarth telescopes observe a star roughly every 20 minutes, while TRAPPIST-UCDTS focuses on one target at a time with a cadence of the order of a minute. We therefore had to find a criterion to find flares and remove these from the light curves.

Using \textit{Kepler} data, \cite{Davenport_2014} generated a median flare template from 885 flares observed on the M4 star GJ 1243 to analyse the different phases of a flare. Their template suggests a flux rise, characterised by a fourth order polynomial, followed by an initial impulsive exponential decay which smoothly transitions into a more gradual exponential decay phase. As other authors \citep[e.g.][]{Walkowicz_2011,Loyd_2014}, we approximated a flare's shape as a single exponential decay. This simple model is sufficient for our purposes since we iteratively fit and remove flares and because a gradual flux decay within the typical flux range is not expected to confound the transit search code.

Flare candidates in the differential target light curves are identified by evaluating criterion~(\ref{eq:flares}) with condition (\ref{eq:condition}) for all flux values:

\begin{equation}
	\: \frac{2 f_{j}-f_{j-2}-f_{j+3}}{\sigma_j} \:\cdot \: \frac{\left|f_{j} - f_{j-2}\right|-\left| f_{j+1} - f_{j}\right|}{\sigma_j}>12
    \label{eq:flares}
\end{equation}
\begin{equation} 
	 2 f_{j}-f_{j-2}-f_{j+3}>0,
    \label{eq:condition}
\end{equation}
where $f_j$ is the $j^{\text{th}}$ flux measurement of the respective star and $\sigma_j$ is the RMS of the 60 data points closest to $j$.

Criterion~(\ref{eq:flares}) can be understood as follows.
For any data point, we compare the potential flux peak $f_j$ to adjacent flux measurements to find flares by exploiting the fact that there is a peak in the light curve and that the flux rises more rapidly than it decays. If there is a flare in the light curve and it peaks at $f_j$, then $f_{j}-f_{j-2}$ captures the flux rise and is positive. In this case, $f_j-f_{j+3}$ is a measure for how quickly the flux decays after the flare peak and is positive as well. The first component of criterion~(\ref{eq:flares}) favours strong signals consisting of a very sharp rise in flux followed by a fast decay since in this case both $f_{j}-f_{j-2}$ and $f_j-f_{j+3}$ are positive and add up.

In the second part of criterion~(\ref{eq:flares}) we exploit the asymmetry of a flare's shape since $f_{j}-f_{j-2}$ is expected to be much greater than $f_j-f_{j+1}$. In this case, $f_j-f_{j+1}$ was chosen over $f_j-f_{j+3}$ since $f_j-f_{j+1}$ captures the asymmetry better. To avoid false signals from noisy data, we divide both parts of criterion~(\ref{eq:flares}) by the sigma clipped standard deviation of the flux differences\footnote{For each data point $j$, we compute the difference between two adjacent flux measurements for the 60 data points closest to $j$. Towards the beginning or end of a night's observation period, the number of included data points gradually shrinks to 30. A flare in the data leads to a high standard deviation of the flux differences, which decreases the strength of the detection signal. Therefore, we clip the flux differences iteratively with $\sigma_{\text{upper}}$ equal to 6 and $\sigma_{\text{lower}}$ equal to 10. A lower sigma leads to a higher number of detections but also to more false positives, which are expected to be filtered out by requiring the flare amplitude to exceed a certain threshold.}. The cut-off value was set to 12 by inspecting the value of criterion~(\ref{eq:flares}) for the smaller flares that we still intend to detect and remove. Assuming the shape of the flare to consist of an abrupt rise followed by an exponential decay, setting the cadence to one minute, the characteristic decay time to 5 minutes, and the standard deviation to 0.5 per cent, we find the value of criterion~(\ref{eq:flares}) exceeds 12 if the ratio of the flare amplitude to the standard deviation is about 3. Preliminary tests with SPECULOOS data suggest that a lower cut-off might be adequate for other data sets (C.A. Murray -- priv. comm.)

Condition~(\ref{eq:condition}) ensures that signals which resemble inverted flares are not removed. We always use $f_{j}-f_{j-2}$ instead of the more intuitive $f_{j}-f_{j-1}$ because the typical exposure time (40 seconds) is relatively long, and thus a flare can begin just shortly before the shutter is closed. This results in a less steep rise in flux in the light curve that does not accurately represent the actual flux evolution of the star. Thus, we avoid the potential distortion of the true flare signal by comparing the potential peak to the penultimate data point before it. Also, it is possible that the flare indeed rises over the duration of two data points, in which case $f_{j}-f_{j-2}$ leads to a stronger signal.

To avoid removing flare-like signals within the typical range of the flux values, we only fit and remove signals if the peak is at least three times the sigma-clipped running RMS above the running median of the data. 

\begin{figure}
    \centering
    \includegraphics[width=.9\columnwidth]{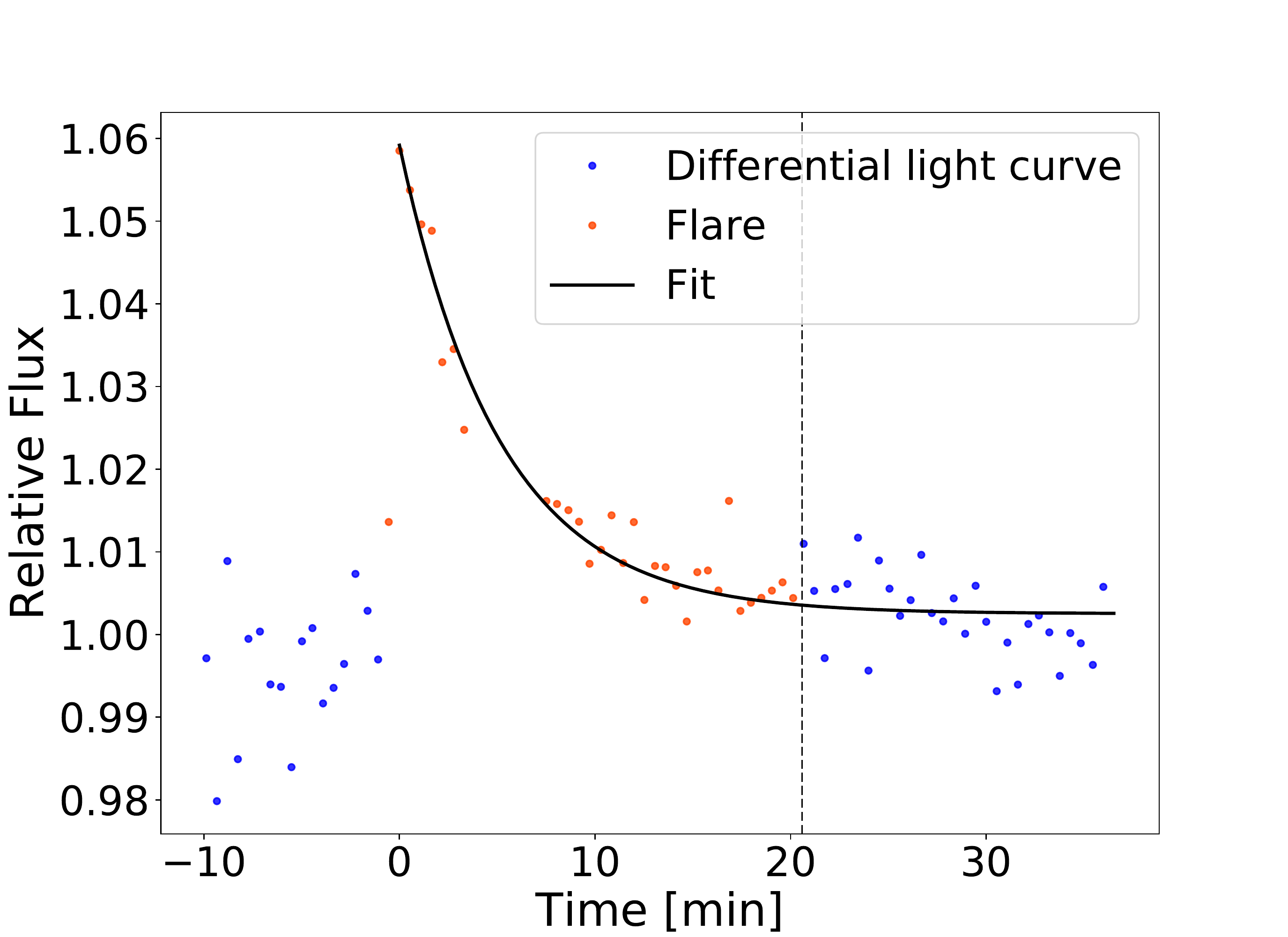} 
    \qquad
    \includegraphics[width=.9\columnwidth]{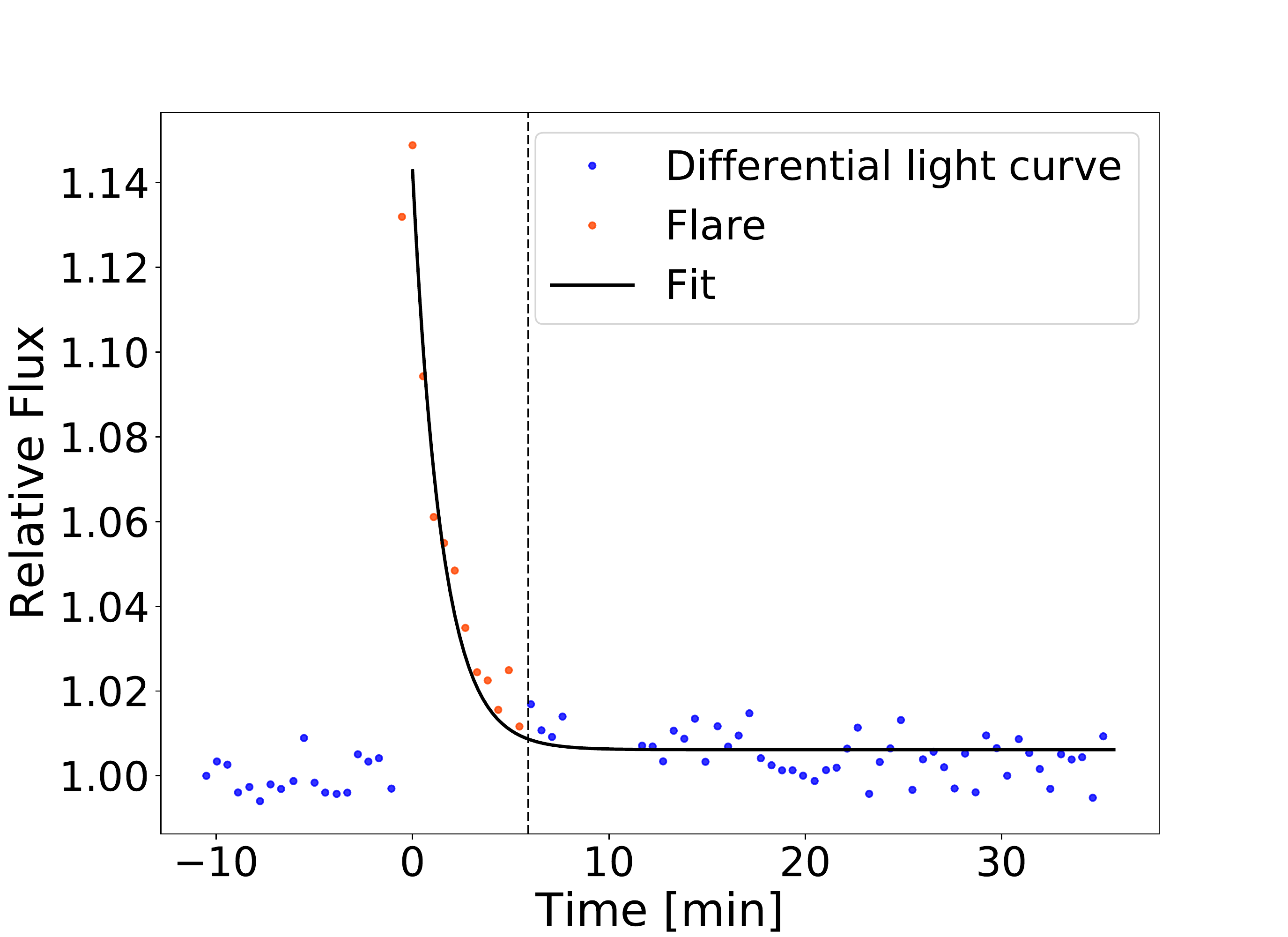} 
    \label{fig:flare1}
    \caption{Two flares in the light curve of LEHPM 2-783 with an exposure time of 21 seconds. In the bottom plot, the flux rises over 2 data points to its maximum, which justifies using $f_{j}-f_{j-2}$ in the flare detection criterion. We remove all flux values from the data point just before the flux peak to the dashed vertical line (in orange).}
\end{figure}

The identified flares are removed by fitting an exponential decay to the data points and removing the data starting from the data point just before the flare peak to four times the characteristic decay time after the flare peak, as shown in Fig. \ref{fig:flare1}. We repeat this procedure until no additional flares are found.

\subsection{Removal of measurements affected by cosmic rays} The electron count within an aperture is significantly increased if a cosmic ray hits a pixel. To circumvent this problem, we discard all images for which the flux of the target star is four times the running RMS above the running mean of its light curve.

\subsection{Light curve detrending from pixel position} The flat-field calibration corrects for pixel sensitivity variations. We refine this calibration and correct residual differential pixel sensitivity variations and potential gaps between pixels, which can affect the light curve if the target moves on the CCD. This effect can be modelled by fitting a 2D second order polynomial as a function of pixel position of the target to the differential target light curve before and separately after the meridian flip of the telescope due to the German equatorial mount. Dividing the target light curve by the fitted polynomial for each observation night before and after the meridian flip removes most of the flux dependence on the pixel position. The standard deviation of the star's CCD position during an observation night, before or after the telescope flip, is typically about 1 pixel. A very similar approach was taken by \cite{Berta_2012} who corrected for correlations between light curve trends and external parameters such as the pixel position in their instrumental systematics model. They also allowed different baselines before and after the meridian flip, which is conceptually the same as our approach. Similar to \cite{Berta_2012}, we do not include a correction for a potential correlation between the FWHM of the target star and its light curve because we find the Pearson correlation coefficient to be usually between -0.1 and 0.1 indicating no photometrically relevant correlation.

\subsection{Light curve detrending using Gaussian Process model}\label{gpmodel} We employ Gaussian Process \citep[GP;][]{gp_2006} regression to model the intrinsic stellar brightness variations, as well as residual instrumental and atmospheric variations not fully accounted for in our initial detrending. The model is intended to capture the general trend over the typical variation timescale, but not short abrupt signals.

We model long-term variations due to effects such as precipitable water vapour \citep{Bailer-Jones_2003, Berta_2012, Murray_2020}, varying target FWHM, and intrinsic flux variations due to starspots or the inhomogeneous cloud coverage of UCDs. Short-term fluctuations in brightness, which can also result from precipitable water vapour changes \citep{Murray_2020}, should remain in the corrected light curves, however.

GP modelling requires an informed choice of kernel covariance function $k$, the mean function $\mu$, and an estimate of the uncertainty of each observation $\sigma_y$. The kernel is used to compute the covariance between every two measurements (Eq. \ref{eq:kernel} in the present analysis), which results in an $n \: \times \:  n$ covariance matrix for $n$ data points. To account for noise in the data, the squared uncertainty of the data points is added on the diagonal.
Combined with the mean function, the covariance matrices then provide a Gaussian distribution at each point of interest.

We use the GP package Celerite \citep{celerite}, which is computationally fast but restricts us in the choice of the kernel. In general, the computational cost of GPs scales cubically with the number of data points. Celerite, however, exhibits linear scaling, which is advantageous for the analysis of a large data set such as ours.

 Several kernels were tested, of which the stochastically-driven damped simple harmonic oscillator (SHO) kernel with the quality factor set to $1/\sqrt{2}$ and the mean function $\mu$ set to the median flux yielded the best results regarding flexibility and applicability to all light curves. This treads an intermediary path between smooth periodic and rough stochastic variations. The kernel of the SHO with the mentioned quality factor is equal to:
\begin{equation}
    k(\tau) = S_{0} \: \omega_{0} \: e^{-\frac{1}{\sqrt{2}} \omega_{0} \tau} \text{cos}\left( \frac{ \omega_{0} \tau }{\sqrt{2}} - \frac{\pi}{4} \right),
    \label{eq:kernel}
 \end{equation}
 where $\tau$ is the time difference between two measurements. The free hyperparameters $S_{0}$ and $\omega_{0}$ are computed by maximising the likelihood of the data given the model using the L-BFGS-B non-linear optimisation routine \citep{Byrd_1995,Zhu_1997}. L-BFGS-B is based on the Broyden-Fletcher-Goldfarb-Shanno (BFGS) algorithm \citep{fletcher1987} and allows the solving of large non-linear optimization problems with bounds using the gradient projection method. We include the entire light curve of each target star for this procedure to capture the global behaviour of the light curve and lower the impact of transits on the GP hyperparameters. The mean of the Gaussian distribution predicted by the GP for each time point then serves as a model for the flux originating from the star.

\begin{figure}
    \centering
    \includegraphics[width=\columnwidth]{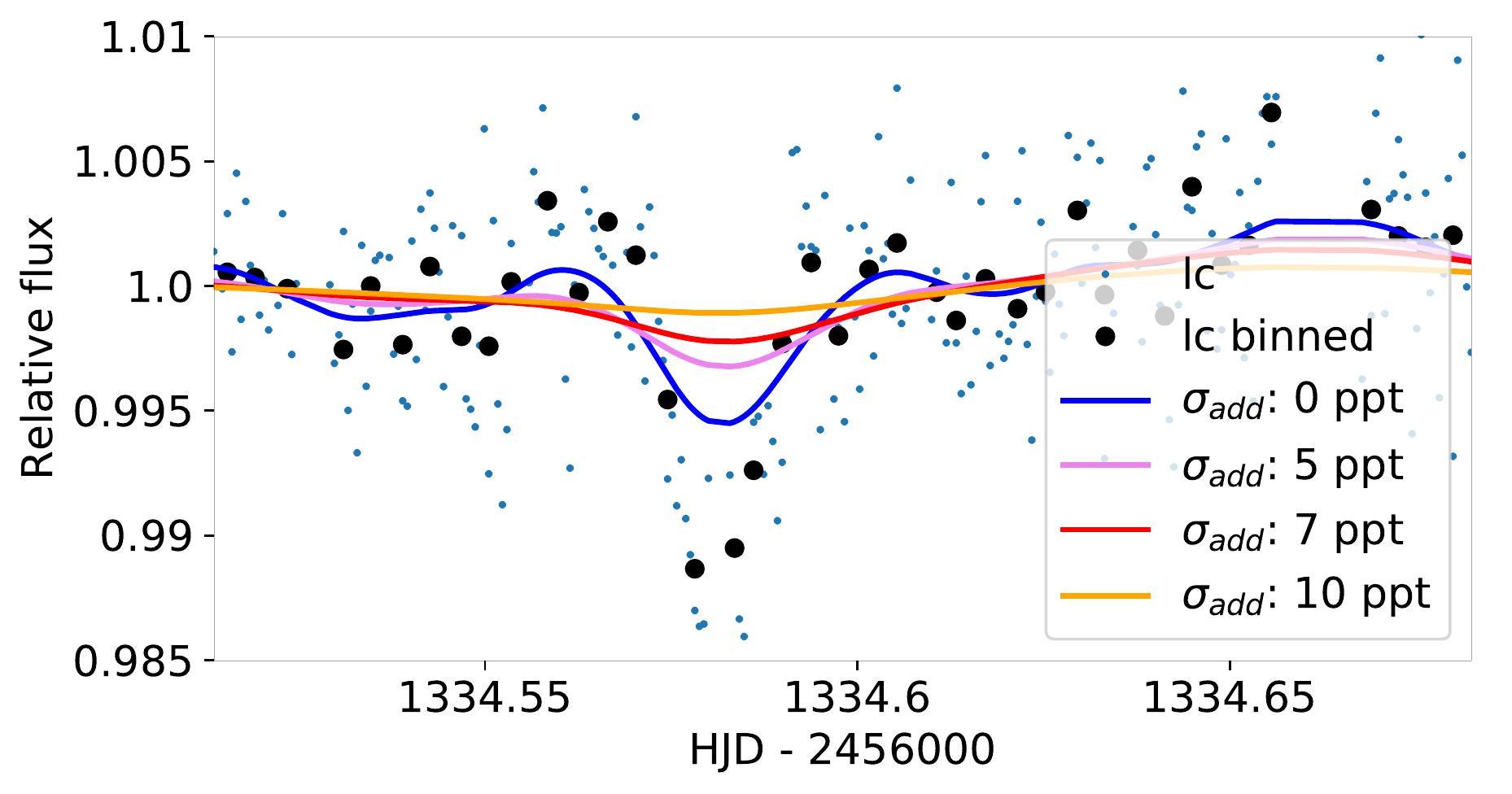}
    \caption{Gaussian Process fit of the normalised differential light curve with a transit of TRAPPIST-1b vs. Heliocentric Julian Date (HJD) applying different measurement uncertainty estimates (blue, violet, red, and orange solid lines). The black dots show the light curve values binned in 7-minute bins. Using the running RMS as the measurement uncertainty leads to a GP mean which closely follows the light curve trends. Quadratically adding a constant additional error term to the running RMS leads to a GP fit that still captures the overall variability of the star itself but less of the planetary transits.}
    \label{gpfit}
\end{figure}

The uncertainty $\sigma_y$ that we need to assign to each data point influences how closely the GP fits the data. We aim at modelling the stellar light curves, but there could be additional planetary transits affecting the measured brightness of their host star. Thus, we set the uncertainty of a flux measurement to the running RMS and add an additional component to the running RMS to reflect this source of uncertainty regarding the true brightness of the star alone as further explained in Section \ref{prepost}. In this analysis, we added 7 ppt to the running RMS ($\sigma_y^2= RMS_{\text{running}}^2 + 0.007^2$) in quadrature, which works well in the case of TRAPPIST-1, for example, as shown in Fig. \ref{gpfit}. 

The running RMS increases at a transit, which additionally lowers the risk of overfitting the light curve and thus hiding transits. To maximise the increase of the uncertainty estimate near a transit, we must make sure that a typical transit fits into the range of data points that we include for the running RMS. For computational efficiency, we set the running RMS of each data point again to the RMS of a fixed number of data points $n_{\text{dp}}$ nearest to the respective data point. Since a transit of a habitable planet orbiting a UCD is expected to take about one hour, we choose $n_{\text{dp}}$ for each star and day such that the median of the time difference between the last and the first included data point is equal to one hour. Transits are not expected to be ubiquitous in the light curve, and therefore we do not expect them to significantly impact the final GP hyperparameters and consequently the estimate of the typical variation time scale.

The FWHM of the kernel function is generally between 1 and 9 hours, while the logarithm of the amplitude hyperparameter varies between -18 and -12.

\begin{figure*}
    \centering
    \includegraphics[width=\textwidth]{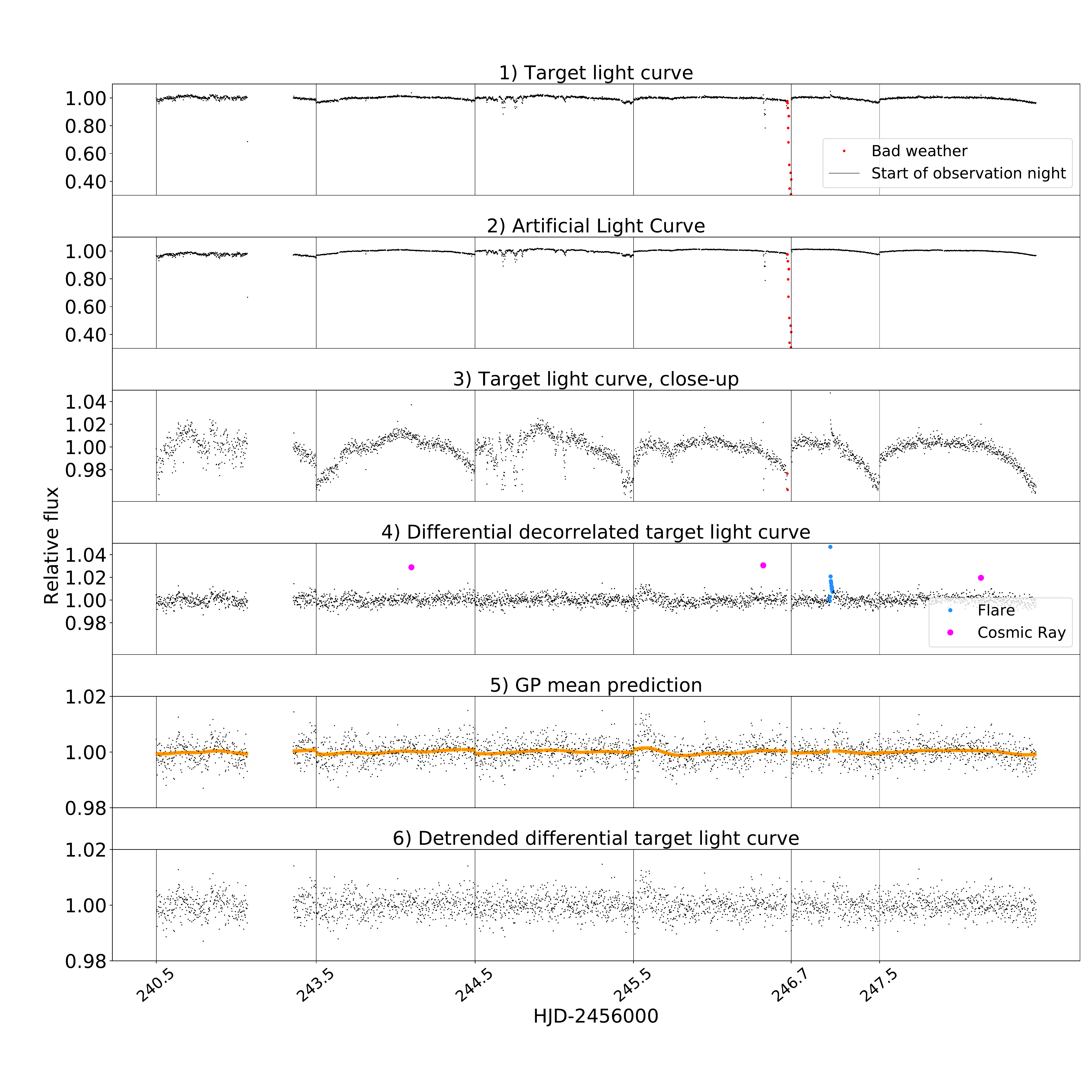}
    \caption{Top to bottom: 1) The light curve of the target star LP  993-98 in Relative flux vs. Heliocentric Julian Date. Observation nights are separated by a black vertical line. Noisy Artificial Light Curve (ALC) parts are displayed in red.
        2) The ALC is the weighted mean of the light curves of the stars surrounding the target star. Noisy ALC parts are displayed in red.
        3) Target light curve as in 1) but on a different scale.
        4) Differential light curve decorrelated from pixel position. Noisy ALC data points have been removed. The flares in the light curve (in blue) as well as data points affected by cosmic rays (in violet) will be removed in the following, resulting in the light curve in panel 5 (LC5).
        5) Gaussian Process mean prediction for LC5, which is very flat in this case.
        6) LC5 divided by the GP mean prediction.}
    \label{star1}
\end{figure*}

\begin{figure*}
    \centering
    \includegraphics[width=\textwidth]{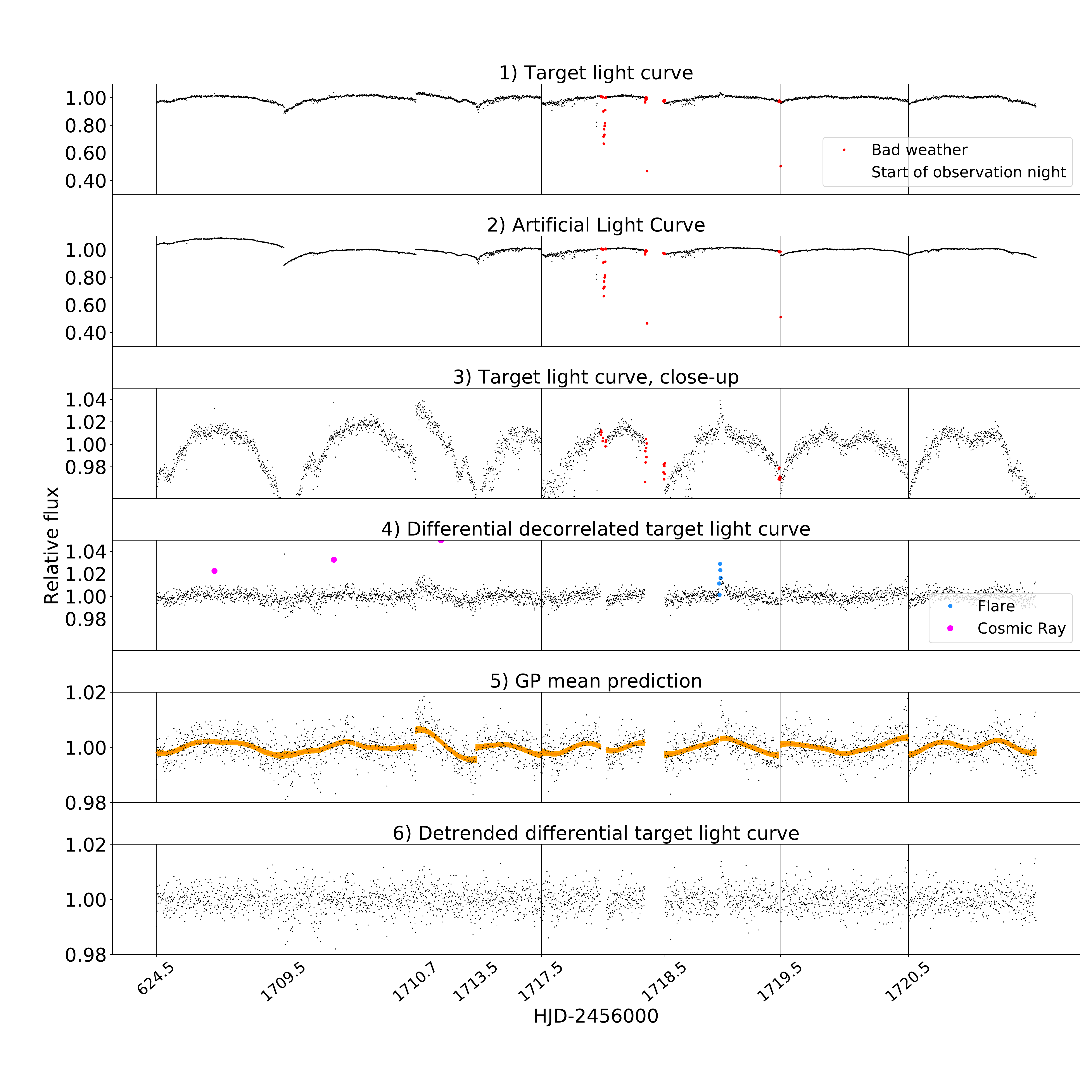}
    \caption{Top to bottom: 1) The light curve of the target star LP  593-68. Observation nights are separated by a black vertical line. Noisy Artificial Light Curve (ALC) parts are displayed in red.
        2) The ALC is the weighted mean of the light curves of the stars surrounding the target star. Noisy ALC parts are displayed in red.
        3) Target light curve as in 1) but on a different scale.
        4) Differential light curve decorrelated from pixel position. Noisy ALC data points have been removed. The flares in the light curve (in blue) as well as data points affected by Cosmic Rays (in violet) will be removed in the following resulting in the light curve in panel 5 (LC5).
        5) Gaussian Process mean prediction for LC5.
        6) LC5 divided by the GP mean prediction.}
        \label{star2}
\end{figure*}

The most important products of the different steps described in this section are displayed for two stars in Figs. \ref{star1} and \ref{star2}.

\subsection{Transit search} To find transits in the GP-detrended differential light curves, we use the Box-fitting Least Squares algorithm (BLS) by \citet{BLS2002}.
BLS folds the light curve on a given number of periods, bins the folded light curves, and fits a box-shaped transit to them. It then provides us with the transit parameters which best fit the analysed light curve. Namely, the period, transit depth, signal residue, and the bin number in the folded light curve where the transit starts and ends respectively. From this, we can reconstruct the position of the BLS-transits in the light curves, and visually compare these potential transits to the neighbouring light curve regions and the ALC to assess the BLS output.

We tested 10,000 orbital periods ranging from 0.8 to 10 days, but no additional planets except those of TRAPPIST-1 were found.

\section{Pipeline output for TRAPPIST-1}  \label{pipeline}
\begin{figure}
    \centering
    \includegraphics[width=\columnwidth]{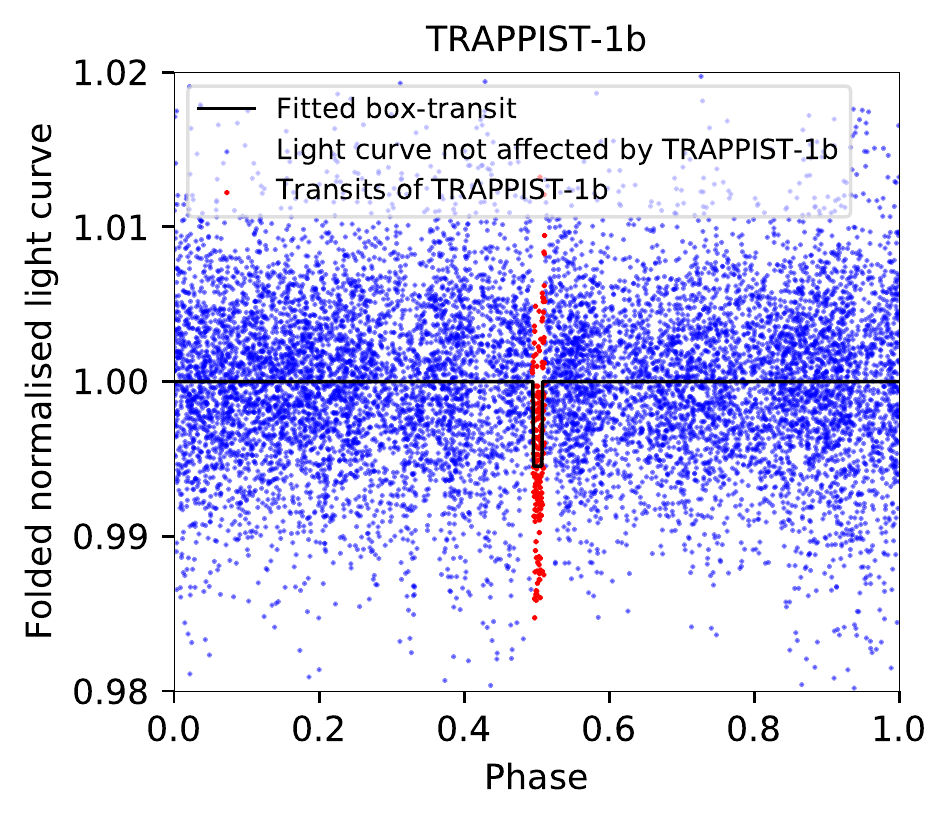}
    \caption{Box Least Squares (BLS) transit for the GP-corrected differential light curve of TRAPPIST-1. BLS correctly identifies all transits of TRAPPIST-1b (in red) and the transit depth. The displayed light curve is folded with the BLS peak period which is equal to the true orbital period of TRAPPIST-1b (1.511 days, \citealt{delrez_trappist_2017}).}
    \label{fig:foldedb}
\end{figure}

\begin{figure}
    \centering
    \includegraphics[width=\columnwidth]{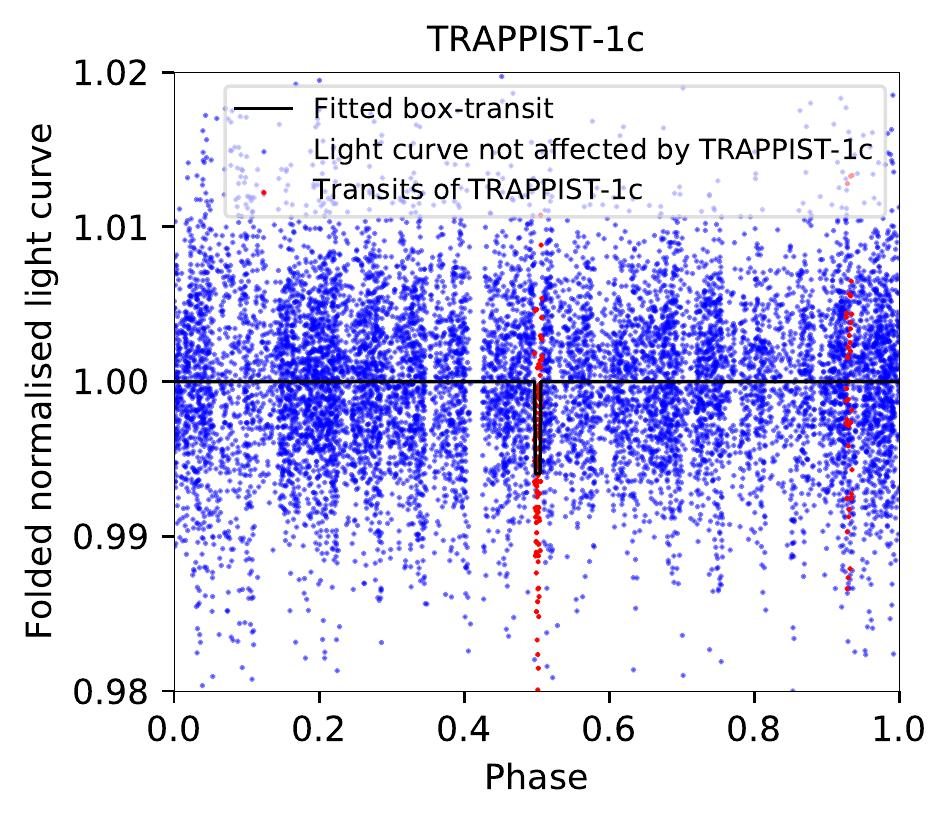}
    \caption{For this figure, we removed the data points affected by the transits of TRAPPIST-1b, recomputed the Gaussian Process fit, and let the Box Least Squares algorithm (BLS) look for transits in the resulting detrended light curve. The light curve is folded with the BLS peak period (3.391 days) which is equal to $\frac{7}{5}$ times the period of TRAPPIST-1c (2.422 days, \citealt{delrez_trappist_2017}). The transits of TRAPPIST-1c are displayed in red. Therefore, the pipeline together with BLS detects TRAPPIST-1c but misses one transit near phase equal 0.93.}
    \label{fig:foldedc}
\end{figure}

We assessed the performance of our pipeline using the data collected on TRAPPIST-1 since it is the only star in the data set known to host transiting planets. Fig. \ref{fig:foldedb} shows the TRAPPIST-1 light curve folded with the BLS peak period and the respective box-shaped transit. It shows that we correctly identify all transits of TRAPPIST-1b (in red) using the completely automated pipeline. We then removed the transits of TRAPPIST-1b from the respective light curve and reran the GP-detrending and BLS. This revealed three out of four transits of TRAPPIST-1c present in the light curve since BLS found an alias of the true period to be the most likely orbital period. Fig. \ref{fig:foldedc} displays the TRAPPIST-1 light curve folded with the new BLS peak period. All but one of the TRAPPIST-1c transits (in red) are located within the region where BLS predicts a transit.

TRAPPIST-1c was transiting on the second observation night while TRAPPIST-1g did so on the third. The pipeline identifies these two transits as originating from the same planet if the raw data of the first three observation nights is fed in. Thus, it would have predicted the planetary system very early on.

To get an idea of the accuracy of the BLS output parameters, we compare our results to the TRAPPIST system parameters as derived from Spitzer data by\\ \cite{gillon_trappist_2017}. 
The transit depth for planets b and c (0.576~per~cent, 0.586~per~cent respectively) are lower than the Spitzer estimates (0.7266$\pm$0.0088~per~cent, 0.687$\pm$0.010~per~cent) while the transit durations for planets b and c (27$\pm$4~min, 44$\pm$12~min) are near the Spitzer values (36.40$\pm$0.17~min, 42.37$\pm$0.22~min).
This is mainly due to the box-shaped transit, which is not an adequate representation of a transit for a transit depth or duration analysis but is sufficient for an efficient transit search. Given this effect, the impact of the GP fitting on the transit depth is expected to be minor for the TRAPPIST-1 light curve. The retrieved periods are both accurate up to less than 2 minutes if we account for the aliasing factor in the period of planet c.

We removed the transits of planets b and c in the TRAPPIST light curve, leaving gaps at the respective positions, and included the resultant light curve in the following analysis.

\section{Flare energies} \label{flareenergies}
The population of early and mid-type M dwarfs hosts a comparably high fraction of flaring stars. In a study by \cite{guenther_flares_2019} based on two months of TESS mission data \citep{Ricker_2014}, about 10 per cent of early M dwarfs and about 30 per cent of mid M dwarfs showed observable flares. The flare statistics for late M dwarfs and brown dwarfs are less well understood. We detected a considerable number of flares in our UCD data set and estimated the flare energies as part of this global analysis.
For this, we followed the approach outlined in \citet{shibayama_2013}, which is based on the assumption that the spectrum of white-light flares is consistent with blackbody emission $B_{\lambda}$ at a constant effective temperature.
We set the effective temperature of the flare $T_{\text{flare}}$ to 9,000~K \citep[as in][]{kretzschmar_2011,guenther_flares_2019}.
Based on the normalised light curve evolution, we know the excess flux $C^{\prime}$ originating from the flare relative to the stellar flux, i.e. the ratio between the observed luminosity of the flare divided by the observed luminosity of the star. Assuming the effective temperature $T_{\text{eff}}$ of the UCDs to be 2,700~K and that the effective temperature of the flare is constant, we computed the area $A_{\text{flare}}$ of the flare taking account of the response function of the TRAPPIST instrument $R_{\lambda}$ and using the stellar radii $R_{\ast}$ in Table \ref{startable}:
\begin{equation}
    A_{\text{flare}}(t) = C^{\prime}(t) \pi R_{\ast}^2 \frac{\int R_{\lambda} B_{\lambda}(T_{\text{eff}}) d\lambda}{\int R_{\lambda} B_{\lambda}(T_{\text{flare}}) d\lambda}.
\label{eq:flarearea}
\end{equation}
The response function $R_{\lambda}$ is equal to the CCD window transmission multiplied by the quantum efficiency, the reflectance of the mirrors (twice) and the transmission function of the I+z$'$ filter. 
The estimate of the flare area then serves to compute the bolometric flare luminosity for each flux measurement by the Stefan-Boltzmann law, whereas the total flare energy is computed by integrating over the duration of the flare:
\begin{equation}
    E_{\text{flare}} = \int_{\text{flare}} \sigma_{\text{SB}} T_{\text{flare}}^4 A_{\text{flare}}(t) dt,
\label{eq:flareenergy}
\end{equation}
where $\sigma_{\text{SB}}$ is the Stefan-Boltzmann constant.
\citet{shibayama_2013} estimated the uncertainty in the flare energy to $\pm$~60 per cent. Since we used ground-based measurements while \citet{shibayama_2013} used \textit{Kepler} data, we expect the uncertainty of our estimates to be higher due to the uncertainty in our ALC solution. We neglected the transmission of the atmosphere in Eq. \ref{eq:flarearea} because it affects the result at a level below 0.1 per cent of the flare energy as computed using ESO's SkyCalc Sky Model Calculator based on \citet{Noll2012} and \citet{Jones2013}.

The results are displayed in Table \ref{restable}. The range of the flare energy estimates is similar to but lower than the values for M dwarfs found by \citet{guenther_flares_2019}. For most targets there is an effective temperature estimate derived by the SPECULOOS team as described in the appendix of \cite{Gillon_2020}. The majority of these effective temperatures is near 2,700~K, the value we used to compute the flare energy estimates. Computing the flare energies with the effective temperature of the stars set to 2,000~K leads to results which are one order of magnitude lower than those in Table \ref{restable}.

\begin{table}
\caption{This table lists the flare energy estimates based on the approach described in \citet{shibayama_2013}. We detected a considerable number of flares but no superflares with energies above $10^{34}$~erg.}
\label{restable}
\begin{tabular}{lll}
\hline
Star id                 & $\log_{10}$(flare energy [erg])       & $\text{T}_{\text{eff}}$  \\
     &       & {[}K{]}  \\

\hline
2MASS J03111547+0106307 & 31.4                               & --   \\
2MASS J03341218-4953322 &                                    & --   \\
2MASS J07235966-8015179 &                                    & 2827 \\
2MASS J08023786-2002254 &                                    & --   \\
2MASS J11592743-5247188 & 30.3, 31.0                         & 2355 \\
2MASS J15072779-2000431 &                                    & --   \\
2MASS J15345704-1418486 &                                    & 2502 \\
2MASS J20392378-2926335 &                                    & 2928 \\
2MASS J21342228-4316102 &                                    & --   \\
2MASS J22135048-6342100 & 30.3, 30.7, 30.7, 30.8             & 2889 \\
APMPM J2330-4737        &                                    & 2738 \\
APMPM J2331-2750        &                                    & 2573 \\
DENIS J051737.7-334903  &                                    & 2656 \\
DENIS J1048.0-3956      &                                    & 2360 \\
GJ   283 B              &                                    & 2803 \\
GJ 644 C                & 30.1, 30.4, 30.9                   & 2674 \\
LEHPM 2- 783            & 30.9, 31.5, 31.5, 31.5, 31.9,      & --   \\
                        & 32.8                               &      \\
LHS  1979               &                                    & --   \\
LHS  5303               & 30.4, 30.6, 30.8                   & 2903 \\
LP  593-68              & 31                                 & 2819 \\
LP  655-48              & 32.1                               & 2714 \\
LP  666-9               & 30.6, 30.9, 31.0, 31.2, 32.7       & 2400 \\
LP  698-2               &                                    & --   \\
LP  760-3               & 30.4                               & 2716 \\
LP  775-31              &                                    & 2863 \\
LP  787-32              &                                    & --   \\
LP  789-23              &                                    & 2778 \\
LP  851-346             & 30.1, 30.5                         & 2687 \\
LP  911-56              & 30.6                               & 2824 \\
LP  914-54              & 31.8                               & 2701 \\
LP  938-71              & 30.5                               & 2648 \\
LP  944-20              &                                    & 2313 \\
LP  993-98              & 31.1                               & --   \\
LP 888-18               &                                    & 2642 \\
SCR J1546-5534          & 30.5, 30.5, 30.8, 31.0             & 2758 \\
SIPS J1309-2330         & 30.9                               & 2647 \\
TRAPPIST-1              & 31.0, 31.1                         & 2629 \\
UCAC4 379-100760        & 30.4, 30.5                         & 2976 \\
V* DY Psc               &                                    & --   \\
VB 10                   &                                    & 2578\\
\hline
\end{tabular}
\end{table}

\section{Transit injection tests}  \label{injectiontests}
To assess the sensitivity of the survey, we calculated the probability of detecting planets orbiting their stars with a given orbital period and radius. This task involves estimating the probability that the planetary orbit allows transits ($\mathbb{P}_{\text{geometry}}$) as well as the probability that we identify a planet in such an orbit ($\mathbb{P}_{\text{identify}}$). 
The former can be computed based on the stellar and tested planetary parameters alone. The latter is estimated by simulating the effect of such planets on the light curves and testing whether we can recover the transits. The planet occurrence rate for UCDs are not well-known and could vary within the group of UCDs. Due to the size of our sample, we treat the UCDs as a homogeneous group in this analysis.

\subsection{Planetary and stellar parameters} \label{radiiandmasses}

We tested the ability of the pipeline to detect planets for different orbital parameters and planetary radii by generating 10,000 sets of orbital period ($P$), phase, inclination ($i$) and planetary radius ($R_p$) for each star. Since there are 40 targets in UCDTS, we evaluated the light curves resulting from 400,000 different configurations in total. In this analysis, we restricted the orbits to circular ones. The period, radius, and phase were randomly drawn from a uniform distribution, whereas for the inclinations $i$ we can exploit the fact that cos($i$) is uniformly distributed for randomly chosen orbits. We did not test orbits with inclinations which lead to impact parameters greater than 1 (i.e. inclination angles smaller than $\frac{\pi}{2} - \mathrm{arctan} (\frac{R_{\ast}}{a})$ with $R_{\ast}$ being the stellar radius and $a$ the semi-major axis) since these transits are unlikely to be recovered by BLS and for an observer they are difficult to distinguish from stellar or atmospheric effects. We account for the orbits with impact parameter above 1 in Section \ref{section_detprobs} with $\mathbb{P}_{\text{geometry}}$.

The \citet{mandel_2002} transit model in PyTransit \citep{parviainen_2015} was used to compute model transit light curves.

In addition to the randomly drawn orbital parameters, PyTransit needs the stellar masses, radii, and limb darkening coefficients.
The limb-darkening coefficients were set to [0.65, 0.28], which are those of TRAPPIST-1 for the I+z$'$ filter in \citet{gillon_trappist_2016} inferred from \citet{Claret_2011}. The extent of limb-darkening varies by stellar host, but all stars are similar to TRAPPIST-1 in spectral type. We found the results to depend only marginally on the limb-darkening coefficients.

The stellar mass was derived as in \citet{benedict_2016} using the apparent 2MASS $K_S$-band magnitude and the Gaia DR2 parallax.
The absolute $K_S$ magnitudes of seven targets are above 10, where the stellar mass estimate of \citet{benedict_2016} is not valid. For these stars the mass estimation polynomial leads to a very low mass estimate. To avoid using non-physical stellar masses, we set a minimum of 0.075~$M_{\odot}$ to the stellar mass estimate. This mass was chosen because it is the approximate mass of an M9 star \citep{reid_2005}. Some of the targets could be brown dwarfs for which we overestimate the mass by setting a minimum of 0.075~$M_{\odot}$. Consequently, this leads to an underestimated probability of finding the planet in a transiting orbit. However, setting a lower limit does not change the results significantly and thus, the results overall do not depend critically on the exact value of the lower limit.
We neglect reddening since all targets are within 30~pc and therefore within the Local Bubble where reddening is negligible \citep{holmberg_2007}.

Next, we needed to estimate the stellar radius to infer the transit depth of a planet with a given size. The relationship between the $K_S$-band magnitude and the radius for M dwarfs described in \citet{mann_2015} provides the desired estimate for $K_S$ below 9.8. Since the radius-magnitude relationship still provides estimates in the expected radius range even if the $K_S$-band magnitude is slightly above 9.8, we used the Mann radius estimate for all target stars. The uncertainties of the Gaia parallax and the $K_S$ magnitude measurements are low and \citet{mann_2015} report a fractional residual of less than 5 per cent for their fitting polynomial. However, we applied their polynomial for stars at the upper end of the absolute magnitudes considered. Therefore, the estimates suffer from low number statistics. 

As the last input parameter of PyTransit, the semi-major axis scaled by the stellar radius was derived using Kepler's third law.

Light curves with simulated planets were then computed by multiplying the real final light curves by the model light curves from PyTransit.
Running the BLS algorithm on the product then provided us with a guess for the location, depth, and duration of the injected transits. Next, we need to define when BLS result is to be considered as a correct recovery.

\subsection{Identification criterion} \label{identificationcriterion}

\begin{figure}
    \centering
    \includegraphics[width=\columnwidth]{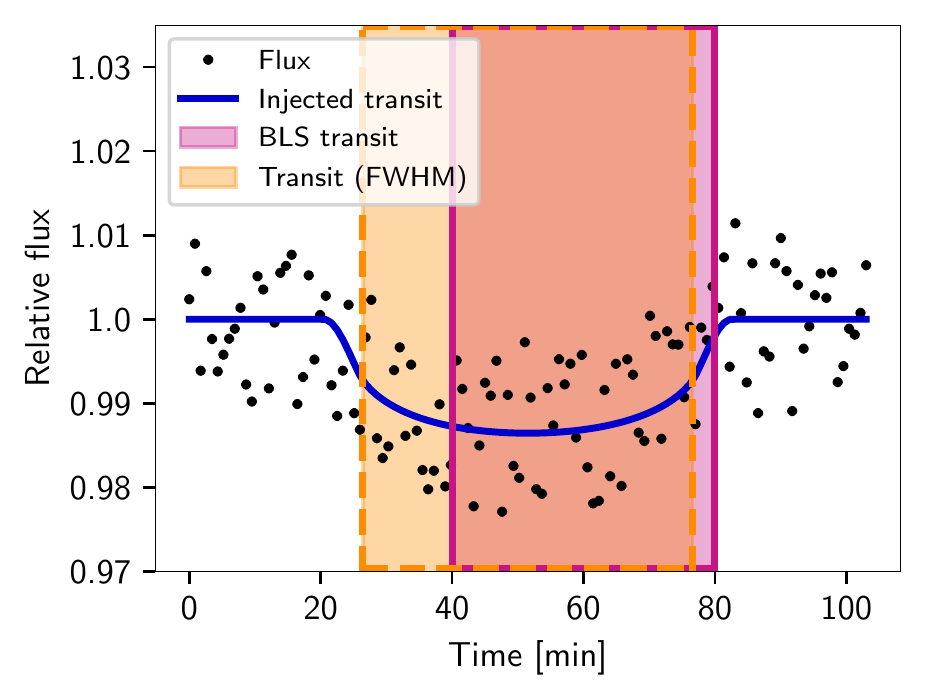}
    \caption{Illustration of the detection criterion. A synthetic light curve was multiplied by a transit light curve generated with PyTransit to display a potential light curve with an injected transit. The area shaded in orange in the dashed line box indicates where the injected transit dips below half of its maximal transit depth. The transits suggested by the Box Least Squares algorithm can be shifted due to noise or other transits in the light curve. We regard a transit as detected if the BLS transit (shaded in violet in the solid line box) overlaps at least half of the transit at half depth (shaded in orange in the dashed line box).}
    \label{detcrit}
\end{figure}

We know at which time point in the light curve we injected transits and where BLS detected a transit-like signal. Thus, we compared the time of the injected transits in the light curve to the BLS-transit region to assess whether we correctly identify the injected planet. Whenever BLS recovered at least half of the transit duration at half of the transit depth (the area shaded in orange in Fig.~\ref{detcrit}), the respective transit is labelled as detected. We count an injected planet as recovered if at least two injected transits were detected by the mentioned criterion.\\ Other authors such as \citet{giacobbe_2012} or \citet{petigura_2013} used a criterion based on the recovered period. We found this period-based criterion to work well for calm light curves. In all other cases it is, however, too restrictive since BLS seeks to include any light curve feature resembling a transit. A transit-like signal in the original light curve can therefore cause the BLS period to differ significantly from the injected period and its aliases even if BLS found all or most injected transits. We assume that the BLS algorithm is used as an initial tool to find potential planetary transits. As done in TRAPPIST-UCDTS, the potential transits can subsequently be vetted by analysing, for instance, the stellar FWHM, the ALC, or the light curves of stars of the same or a similar spectral type. In other studies \citep[e.g.][]{Berta_2013, He_2017}, a planet was counted as recovered if a single transit with detection significance above a certain threshold was found. \cite{Berta_2012,Berta_2013} additionally propagated the uncertainties associated with their detrending pipeline into the detection significance, which is favourable for MEarth, a multi-telescope survey targeting a high number of stars. TRAPPIST-UCDTS consists of a significantly lower number of stars but operates at a higher cadence than MEarth, which enables rigorous vetting of the transit candidates. Furthermore, the low number of stars allows us to keep track of the impact of the detrending procedure. If we can correctly identify two transits originating from the same planet, we can schedule follow-up observations to observe a third transit. For these reasons, the two-transit detection criterion was preferred.

Applying our identification criterion to each BLS result, we counted the number of correctly recovered planets in a given radius and period range to assess the probability ($\mathbb{P}_{\text{identify}}$) to identify such planets. Additionally, we calculated the results for $\mathbb{P}_{\text{geometry}}$ in the same radius and period range and combined the two probability estimates to get the detection probability.

\subsection{Detection probability} \label{section_detprobs}

The probability of detecting a planet in a transit survey depends on the probability of the existence of such a planet, the probability of the planet orbiting its host star such that it periodically transits its star as seen from Earth, and the probability that the observer identifies the signal in the light curve.
The probability of finding a planet orbiting star $j$ is therefore equal to:

\begin{equation}
    \mathbb{P}_{\text{detect}_j}=\mathbb{P}_{\text{planet}} \cdot \mathbb{P}_{\text{geometry}_j} \cdot \mathbb{P}_{\text{identify}_j} \:
\label{pdetgen}
\end{equation}

with

\begin{enumerate}[leftmargin=*,labelindent=0pt,label=$\cdot$]
    \item $\mathbb{P}_{\text{planet}}$: Probability that the tested planet orbits the respective star. This is set to 1 except where stated otherwise.\\
     \item $\mathbb{P}_{\text{geometry}_j}(R_{\ast_j},M_{\ast_j},P)$: Probability that the inclination of the planetary orbit lies within the accepted inclination angle range (impact parameter smaller than 1). Using Kepler's third law, we get the semi-major axis $a$ from the orbital period and the stellar mass. The geometric probability for an orbit with impact parameter below 1 is equal to $\frac{R_{\ast}}{a}$.\\
    \item $\mathbb{P}_{\text{identify}_j}(R_{\ast_j},M_{\ast_j},R_p,P)$: Probability of identifying the planetary transits (cf. Section \ref{identificationcriterion}) if the planet orbits within the accepted inclination angle range.
\end{enumerate}

\subsection{Effect of GP-detrending on identification probability} \label{prepost}

The identification probability mainly depends on how we detrend the light curves before or while searching for transits. In the following, we discuss why we detrend the light curves using a GP before injecting transits and looking for them with BLS (detrend-inject-recovery test).
This approach is intended to be as close as possible to the actual planet detection probability in a survey within which we can analyse light curves in detail and we can apply more sophisticated techniques. The approach taken might appear optimistic. However, a more optimistic detrending approach will lead to a more conservative lower limit on the occurrence rate of planets like TRAPPIST-1b later on. Furthermore, we investigated the effect of the detrending on our occurrence rate estimates to assess the impact on our conclusions.

Injecting transits first, GP-detrending after that and subsequently looking for transits (inject-detrend-recovery test) is not optimal since it necessarily decreases the transit depth. Inspecting a light curve by eye as done in TRAPPIST-UCDTS would likely yield a higher detection probability, and hence inject-detrend-recovery testing underestimates the actual probability of spotting transits. Additionally, detrending after every transit injection is computationally expensive. Detrending and searching for transits simultaneously would be optimal but is even more computationally expensive and therefore not suited for our statistical analysis.

\begin{figure}
	\includegraphics{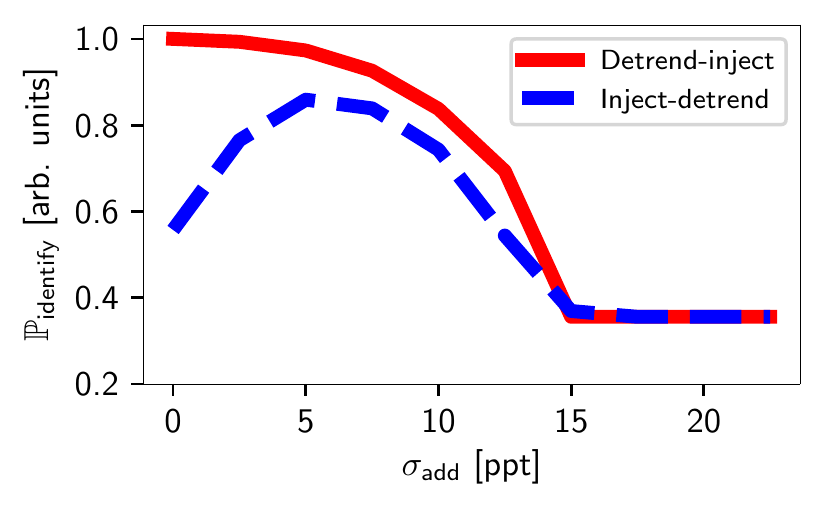}
    \caption{Identification probabilities as a function of the quadratically added uncertainty term $\sigma_{\text{add}}$ for a photometrically variable target with an injected planet with period between 1.4 and 1.8 days and radius between 1 and 1.3 $R_{\oplus}$. Detrending before injecting the transits increases the identification probability estimate. The effect is more pronounced as $\sigma_{\text{add}}$ decreases since a GP fit with a lower $\sigma_{\text{add}}$ produces flatter light curves. Detrending after injecting transits (blue dashed line) can remove a part of the transit but quadratically adding an additional error term $\sigma_{\text{add}}$ mitigates this problem. If we add a very high $\sigma_{\text{add}}$, the GP mean prediction is flat, which is equivalent to not detrending. As visible in this figure, the GP-detrending increases the identification probability in both cases.}
    \label{fig:aderr}
\end{figure}

One necessary component for computing the GP hyperparameters is the uncertainty estimate of the flux measurements. By setting this to the running RMS of the light curve and adding an additional term $\sigma_{\text{add}}$  to it in quadrature, we can adjust how closely we fit the data using the GP, as visible in Figs. \ref{gpfit} and \ref{fig:aderr}.

The GP mean for flat light curves is also flat. This means that the detrend-inject-recovery tests yield a very similar detection probability as a simultaneous approach irrespective of $\sigma_{\text{add}}$. Therefore, we do not need to optimise $\sigma_{\text{add}}$ for these light curves. Instead, we perform inject-detrend-recovery tests on a set of 16 variable light curves with planets similar to TRAPPIST-1b and determine the $\sigma_{\text{add}}$ that maximises the detection probability. This gives us an optimal $\sigma_{\text{add}}$ for each star. Using a GP with $\sigma_{\text{add}}$ set to the mean of these 16 optimised $\sigma_{\text{add}}$, which is equal to 7 ppt, ensures that we mainly remove the general trend and not small transit-like features. We then apply this GP trained with inject-detrend-recovery tests for our detrend-inject-recovery tests. By this procedure, we seek to emulate a simultaneous search.

The impact of the GP-detrending depends on the depth of the transits since the identification probability decays very quickly with transit depth. Therefore, the detection probability of small planets is more prone to being affected by GP-detrending. For a typical variable star, the identification probability for the two test cases is compared in Fig. \ref{fig:aderr} if we inject planets similar to TRAPPIST-1b. The transit detection efficiency over a larger parameter space is compared in Table \ref{comptab} for two stars, which shows that the detection probabilities of the two cases overall converge as we add an additional error term to the running RMS.
In Table \ref{comptab}, we can see again that for the detrend-inject-recovery test case, the identification probability decreases with increasing $\sigma_{\text{add}}$ since a lower $\sigma_{\text{add}}$ leads to a flatter light curve. Transits injected into these flatter light curves are easier to detect. In the inject-detrend-recovery test case, the identification probability increases with increasing $\sigma_{\text{add}}$ because we remove less of the transit depth (cf. Fig. \ref{gpfit}).

In Section \ref{expnr}, we derive a lower limit on the occurrence rate of planets similar to TRAPPIST-1b. As outlined above, this value depends on the detrending. Therefore, we we compute it for different detrending scenarios.

\subsection{Expected number of planets in the data set}
\label{expnr1}

\begin{figure}
	\includegraphics[width=\columnwidth]{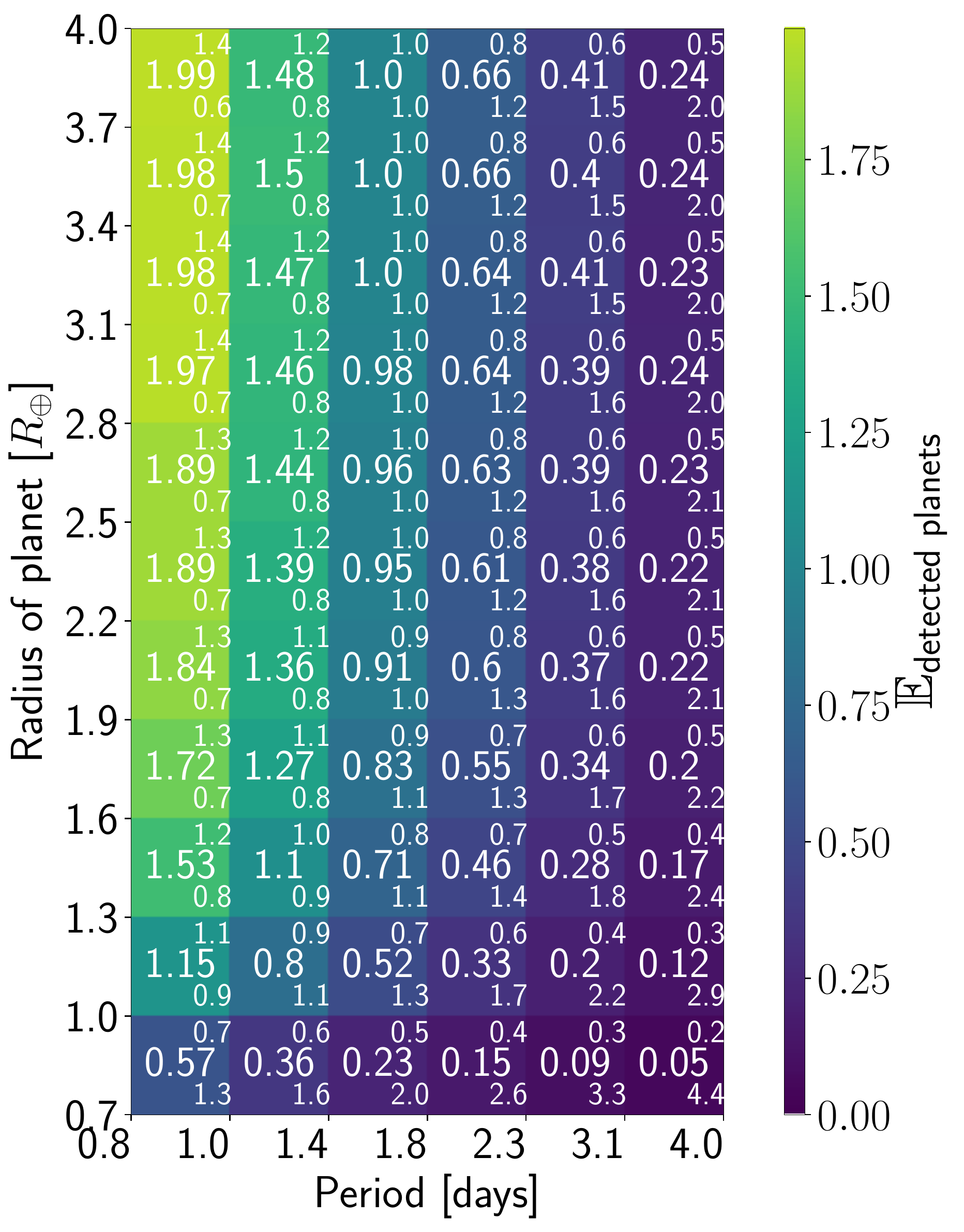}
    \caption{The mean of the Poisson binomial distribution (bold in the centre of each box) is equal to the number of planets we expect to find in the TRAPPIST-South survey (cf. equation \ref{eq:expectationvalue}) assuming an occurrence rate of 1 in each box. The standard deviation of the distribution is shown in the top right corner of each box, while the bottom right number is equal to the skewness of the distribution.
    The mean clearly decreases as a function of period, which is caused by the lower number of transits and the lower geometric probability. The expectation value for the number of planets in the survey saturates above 2 $R_{\oplus}$. Furthermore, we do not expect to detect many planets smaller than Earth. The skewness of the distribution increases as the mean decreases, putting the value for the standard deviation in context.}
    \label{fig:expmap1}
\end{figure}

The probability of finding $n$ planets in the entire sample is described by a Poisson binomial distribution. The mean of this distribution is equal to the number of planets which we expect to find in the data set, defined as
\begin{equation}
	\mathbb{E}_{\text{detected planets}}(R_p,P) = \sum_{j=1}^{\text{nr of target stars}} \mathbb{P}_{\text{detect}_j}(R_p,P),
    \label{eq:expectationvalue}
\end{equation}

in which $R_p$ is the radius of the planet, P is its orbital period, and $\mathbb{P}_{\text{detect}_j}$ is computed as described in Eq. \ref{pdetgen}. In Fig. \ref{fig:expmap1} we illustrate $\mathbb{E}_{\text{detected planets}}(R_p,P)$ with the occurrence rate of the tested planets set to 1.

\begin{figure}
	\includegraphics[width=\columnwidth]{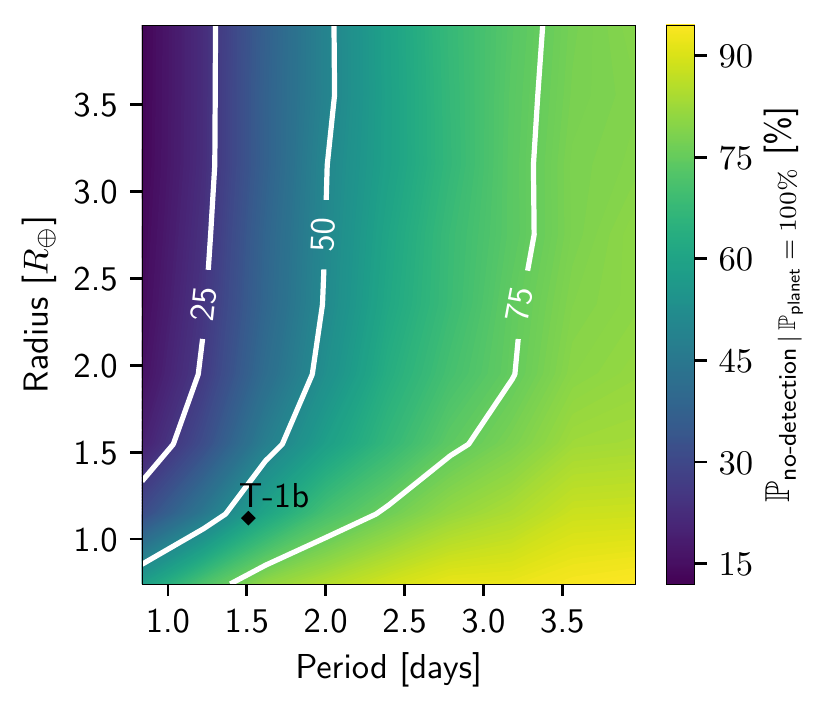}
    \caption{Probability of no single planet detection assuming that all stars host a planet of the tested type in a randomly chosen orbit (cf. equation \ref{eq:nodetectionprobability}). This probability increases sharply for planets with radii below 1 $R_{\oplus}$. Due to the geometric probability decreasing as a function of the period and the lower number of transits in the data, the total no-detection probability also increases for longer orbital periods.}
    \label{fig:expmap2}
\end{figure}

Additionally, we can calculate the probability of not finding any planet for a given star and the probability of not detecting a single planet of a given type in the combined data set (Fig. \ref{fig:expmap2}) using

\begin{equation}
	\mathbb{P}_{\text{no-detection}}(R_p,P) = \prod_{j=1}^{\text{nr of target stars}} \left( 1-\mathbb{P}_{\text{detect}_j}(R_p,P) \right).
    \label{eq:nodetectionprobability}
\end{equation}
As visible in Fig. \ref{fig:expmap2}, an Earth-sized planet with an orbital period above 3 days is unlikely to be detected even if these planets are assumed to be frequent. Furthermore, we can see that the detection of at least one close-in planet in the data set is likely if there is one around every single star with randomly drawn orbital parameters (such as the inclination).
\begin{table}
\caption{Averaged identification probability for two target stars (the same stars as in Figs. \ref{star1} and \ref{star2}) with the period and radius drawn from a uniform distribution between 0.8 and 4 days and between 0.7 and 4 $R_{\oplus}$ respectively. We tested how the quadratically added uncertainty term $\sigma_{\text{add}}$ influences the identification probability as we apply the GP-detrending before we inject transits (detrend-inject-recovery test) or after we have injected transits into the light curve (inject-detrend-recovery test). The results for the two versions differ significantly if we set the uncertainty to the running RMS, but the values converge as we quadratically add a higher additional error term. This indicates that the light curves are not significantly overfitted overall in the detrend-inject-recovery case analysed here.}
\label{comptab}
\begin{tabular}{ |p{3cm}|p{2cm}|p{2cm}|  }
\hline
\multicolumn{3}{|c|}{Identification probability for LP  993-98} \\
\hline
$\sigma_{\text{add}}$& Detrend-inject & Inject-detrend \\
\hline
0 ppt & 19.3\% & 14.8\% \\
5 ppt & 17.9\%   & 15.8\% \\
7 ppt  & 16.7\% & 15.4\% \\
10 ppt   & 16.0\% & 16.0\%\\
\hline
\end{tabular}
\begin{tabular}{ |p{3cm}|p{2cm}|p{2cm}|  }
\hline
\multicolumn{3}{|c|}{Identification probability for LP  593-68} \\
\hline
$\sigma_{\text{add}}$& Detrend-inject & Inject-detrend \\
\hline
0 ppt & 25.3\% & 21.5\% \\
5 ppt & 24.2\%   & 22.1\% \\
7 ppt  & 24.0\% & 22.3\% \\
10 ppt   & 23.6\% & 22.3\%\\
\hline
\end{tabular}
\end{table}

\section{Injection results}
\label{expnr}

\begin{figure}
	\includegraphics[width=\columnwidth]{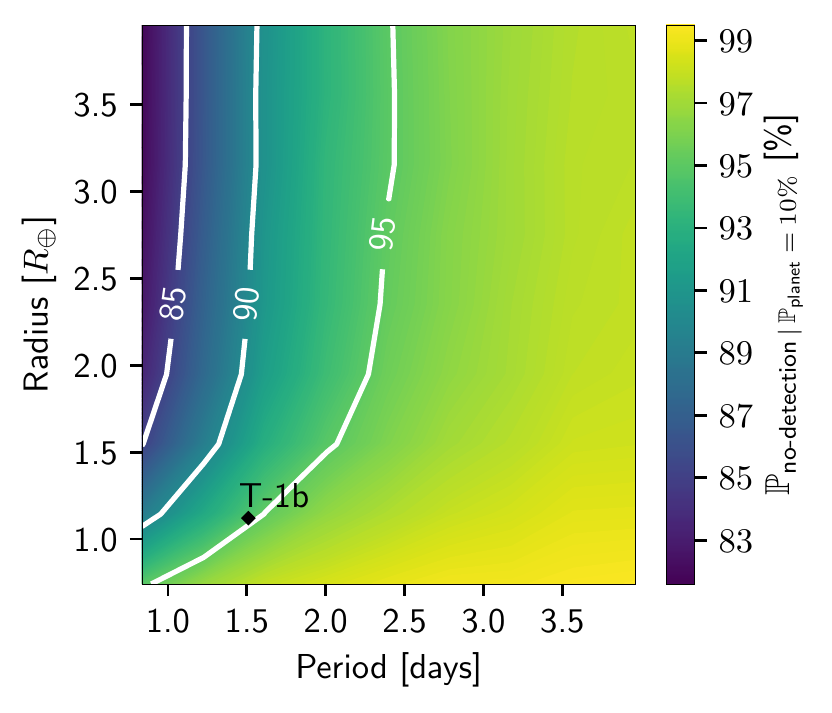}
    \caption{Probability of no planet detection assuming that 10 per cent of all stars have a planet of the tested type. The black diamond shows the period and radius of TRAPPIST-1b.}
    \label{fig:expmap3}
\end{figure}

In Fig. \ref{fig:expmap3} the probability of not detecting a single planet, assuming that 10 per cent of all stars host a planet of the tested type in a randomly chosen orbit, is displayed. This analysis indicates that the probability of finding at least one planet similar to TRAPPIST-1b (in radius and orbital period) in the entire data set is equal to 5 per cent if the occurrence rate for such planets is equal to 10 per cent. In other words, in this case the probability of not detecting a planet like TRAPPIST-1b is equal to 95 per cent and the TRAPPIST team would been very lucky to find a planet as they did. Since TRAPPIST-1b was found in this survey, we can conclude that the occurrence rate is likely to be above 10 per cent.

\begin{figure}
	\includegraphics[width=0.9\columnwidth]{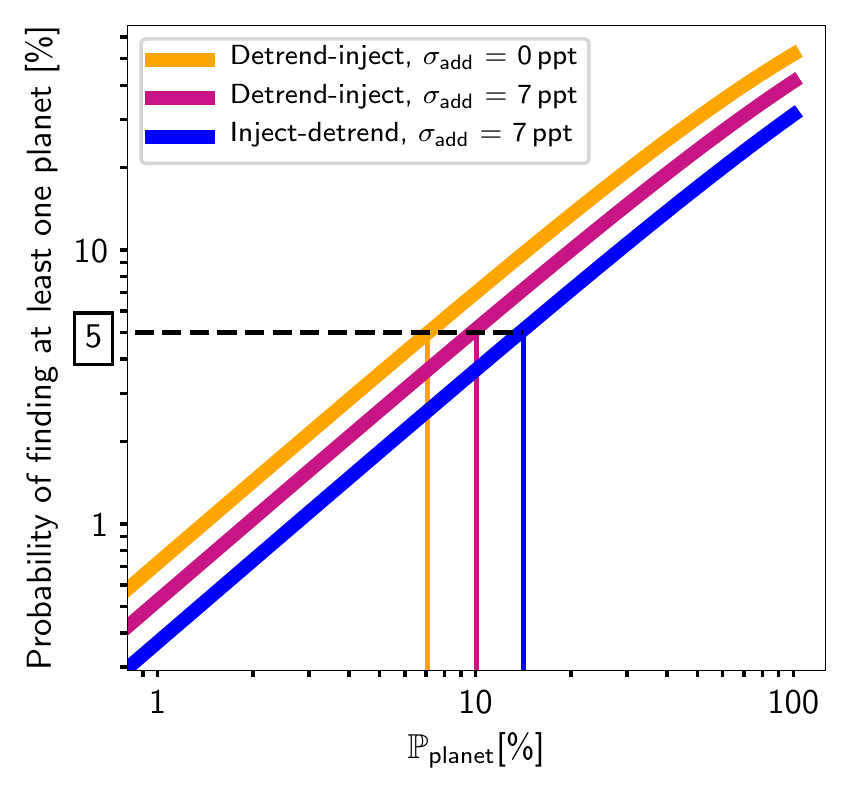}
    \caption{Probability of finding at least one planet similar to TRAPPIST-1b as a function of the occurrence rate resulting from Eq. (\ref{eq:minprob}) for three detrending scenarios. The optimistic scenario (orange) describes the detection probabilities if we detrend the light curves without removing any part of the transits. In the intermediate scenario (violet-red), we still correct for variability without removing transits but we compromise on how well the stellar variability is removed. In the pessimistic scenario (blue), we used the probabilities to identify planets in light curves which have been GP-detrended after the transit injection.
    In the intermediate scenario, if the occurrence rate is below 10 per cent (14 per cent in the pessimistic scenario, 7 per cent in the optimistic scenario), the probability of not finding any planet similar to TRAPPIST-1b in period and radius orbiting a ultra-cool dwarf, exceeds 95 per cent, which means that the probability of finding at least one planet is below 5 per cent.}
    \label{fig:probatleastone}
\end{figure}

In Fig. \ref{fig:probatleastone}, we display the probability of finding at least one planet in the data set as a function of the occurrence rate for three different detrending scenarios. The results quoted here refer to the case where we detrend before injecting planets and quadratically add 7~ppt to the initial uncertainty estimate except where stated otherwise. 

Additionally to the result that the occurrence rate is likely to be above 10 per cent if the likelihood of finding at least one planet is expected to be greater than 5 per cent, we see that an occurrence rate of only 2 per cent would lead to a likelihood of 99 per cent of not detecting a planet like TRAPPIST-1b in such a survey.

\subsection{Dependence on detrending}
As a first estimate of $\mathbb{P}_{\text{planet}_\text{lim}}$, the lower limit on the occurrence rate of planets like TRAPPIST-1b, we derive the occurrence rate for which the probability of no planet detection in the data set is equal to 95 per cent:
\begin{equation} \label{eq:minprob}
 0.95 = \prod_{j=1}^{\text{nr of target stars}} \left( 1- \mathbb{P}_{\text{planet}_{\text{lim}}}\cdot \mathbb{P}_{\text{geometry}_j} \cdot \mathbb{P}_{\text{identify}_j} \right).
\end{equation}

We estimate that the geometric probability is accurate up to 10 per cent, in which case the identification probability dominates the result.

If we overfit the light curve in the GP-fitting or any of the previous steps, we produce very flat light curves, within which injected transits are easy to identify compared to transits injected in more variable light curves. Thus, overfitting before the planet injection leads to a higher identification probability $\mathbb{P}_{\text{identify}}$. 

As visible in equation (\ref{eq:minprob}), the lower limit on the occurrence rate $ \mathbb{P}_{\text{planet}_{\text{lim}}}$ decreases as $\mathbb{P}_{\text{identify}}$ increases.
Consequently, choosing a more optimistic detrending leads to a more conservative estimate of $\mathbb{P}_{\text{planet}_\text{lim}}$.

Ideally, we would search for transits while simultaneously fitting stellar variability.
Finding an efficient algorithm that performs light curve detrending simultaneously with the transit search, and is suited for statistical analyses, is out of scope of this paper. To test how different non-simultaneous detrending scenarios affect the lower limit, we computed it for three different cases.

Using a GP to detrend without simultaneously fitting for transits decreases the detection probability since the GP will try to fit the injected transits. Thus, dividing by the GP mean prediction reduces the depth of the injected transits. We might have found transiting planets by visual inspection that are hidden after GP-detrending. Thus, GP-detrending after the planet injection and before the transit search constitutes the pessimistic scenario. To avoid overfitting the light curves and thus improve the identification probability after detrending, we added an additional component to the uncertainty of each data point in the GP computation. More specifically, we set the uncertainty to $\sigma_y= \left(RMS_{\text{running}}^2 + \sigma_{\text{add}}^2 \right)^{\frac{1}{2}}$ and set $\sigma_{\text{add}}$ to 7~ppt as explained in Section \ref{prepost}.

For the intermediate scenario, we injected transits after GP detrending. This is equivalent to assuming that we can detrend the light curves without removing transits. However, we compromise on how closely the GP fits the data and thus how well it removes stellar and instrumental variability by using the same uncertainty estimate as in the pessimistic scenario.

For the optimistic scenario, the light curves were detrended before transit injection as well. The additional error component $\sigma_{\text{add}}$, however, was set to zero. Consequently, we detrended without removing transits or compromising on the GP fit. This case constitutes the optimistic scenario as simultaneous transit and variability fitting using a GP with an SHO kernel is unlikely to yield a higher identification probability. Therefore, this approach leads to the lowest estimate of $\mathbb{P}_{\text{planet}_\text{lim}}$.
As depicted in Fig. \ref{fig:probatleastone}, the lower limit estimate on the occurrence rate of planets similar to TRAPPIST-1b orbiting UCDs does not change drastically but stays within a range of 7 to 14 per cent depending on whether we detrend before or after injecting planets.

\section{Likelihood analysis}
\label{expnr2}
Alternatively, we can calculate the likelihood of the occurrence rate $r$. $\mathbb{P}(D \: | \: r)$ is the probability of finding a data set $D$ with exactly one planet similar to TRAPPIST-1b around TRAPPIST-1 and no planets around the other surveyed stars, given an occurrence rate $r$ within the parameter range 0.9 to 3 days and 1 to 4~$R_{\oplus}$. This accompanies the probability of finding a planet around at least one of the surveyed stars as described in section \ref{expnr}. We outline the derivation below, while a detailed explanation can be found in Appendix~\ref{sec:occurrence}.

\begin{figure}
	\includegraphics{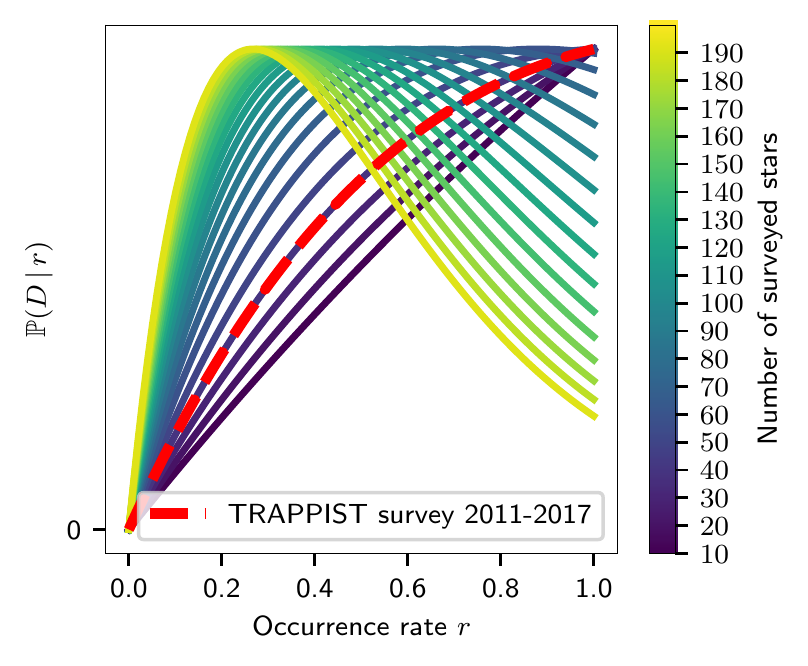}
    \caption{Probability of the survey result as a function of the occurrence rate $r$ for different numbers of surveyed stars divided by the maximum of the respective probability curve. A uniform prior was used to marginalise over the unknown planetary parameters. The most likely occurrence rate for TRAPPIST-UCDTS is currently equal to 1 but shrinks to 0.3 as we observe 200 target stars without detecting another planetary system.}
    \label{fig:pdr}
\end{figure}

\begin{figure}
	\includegraphics{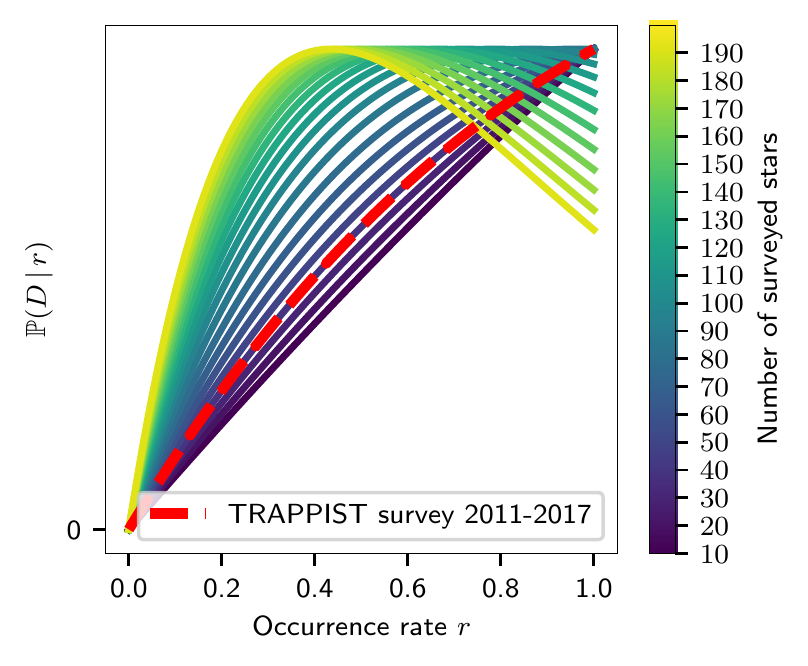}
    \caption{Probability of the survey result as a function of the occurrence rate $r$ except that the normalised Kepler occurrence rates for M dwarfs were used as the prior. This prior essentially excludes planets with orbits below 2 days and radii above 2 $R_{\oplus}$.}
    \label{fig:pdrkepler}
\end{figure}

First, we marginalised the detection probabilities over the analysed radius and period range (0.9 to 3 days, 1 to 4 $R_{\oplus}$) because the planetary and orbital parameters of a potential undiscovered planet are unknown. For this, it is necessary to choose a prior reflecting our knowledge about where in the parameter space planets are likely to be found. Since this is unknown for UCDs, we tested two different priors. As our first prior, we chose a uniform prior. By using this prior for the marginalisation of the detection probabilities, equal weight is given to all potential planets within the tested parameter range. This prior may not be optimal given that for low-mass stars e.g. hot Neptunes are more rare than their warm counterparts \citep[e.g.][]{dressing_2015,Hirano_2018}. As the second prior we chose the normalised occurrence rates of M dwarfs as computed by \cite{dressing_2015} using \textit{Kepler} data. The following steps were carried out separately for the two sets of marginalised detection probabilities resulting from the two priors. The different results are shown in Figs. \ref{fig:pdr} and \ref{fig:pdrkepler}.

We drop the dependence on stellar radii and masses and denote the marginalised detection probability for star $j$ as $F_j$.
The probability of finding a planet around star $j$ results as $r \cdot F_j$, where $r$ is the number of planets we expect to find within the mentioned radius and period range.

Adding over all possible combinations of single-planet systems and weighting them with their corresponding probability, we finally get:
\begin{equation}
    \mathbb{P}(D \: | \: r) = 
    r f(\Theta_{T1b}) \cdot \prod_{j \: \in \: \text{S}} 
    \left( 1-r F_j \right) .
\end{equation}
where $f(\Theta_{T1b})$ is the probability of finding TRAPPIST-1b around its host star and $S$ is the set of all target stars except TRAPPIST-1.
In fact, the detection probabilities for the injected planets also depend on the probability that a human confirms a correct BLS planet detection. This varies depending on the stellar and planetary parameters and the individual observer, but a characterisation of this function is out of scope of this paper. As explained in Section \ref{identificationcriterion}, we assume that the flexible design of the survey and the definition of the identification criterion allow us to set the confirmation probability of a correct BLS detection to $1$.

In Fig. \ref{fig:pdr} and \ref{fig:pdrkepler}, we show $\mathbb{P}(D \: | \: r)$ of TRAPPIST-UCDTS (dashed red line). Furthermore, we averaged over the detection probabilities to further estimate the shape of $\mathbb{P}(D \: | \: r)$ for a higher number of surveyed stars. Since the \textit{Kepler}-informed prior puts a low weight on some regions of the parameter space where the detection probability is high, the marginalised detection probabilities are lower compared to the results derived using a uniform prior. Therefore, the peak of $\mathbb{P}(D \: | \: r)$ is at a higher $r$ in Fig. \ref{fig:pdrkepler} than in Fig. \ref{fig:pdr} for the same number of stars.

The low number of target stars restricts us from drawing strong conclusions on the occurrence rate of short-period planets hosted by UCDs. A larger sample would be more indicative, as illustrated in Fig. \ref{fig:pdr}.

From a Bayesian perspective \citep[e.g.][]{Hall_2018}, if one chooses a prior distribution on $r$, then the likelihood $\mathbb{P}(D|r)$ may be inverted to a posterior distribution $\mathbb{P}(r|D)$. If the prior is uniform $\mathbb{P}(r)=1$ then $\mathbb{P}(r|D)\propto \mathbb{P}(D|r)$, and the curves in Fig. \ref{fig:pdr} may be interpreted as Bayesian degrees of belief in the unknown parameter $r$.

\section{Conclusions}  \label{conclusion}

\subsection{Survey results}
We performed aperture photometry, generated differential light curves and cleaned these from non-planetary signals (cf. Section \ref{Methods}) and also from slow stellar and residual instrumental variability (cf. Sections \ref{gpmodel}, \ref{prepost}). Through injection-recovery tests we could then infer, inter alia, the number of planets we expect to find in such a survey (cf. Section \ref{expnr}). The results rely on our estimates of the stellar parameters in Table \ref{startable} and the applied identification criterion in Section \ref{identificationcriterion}.
We expect to find 0.52 planets similar to TRAPPIST-1b in radius and period and 0.2 planets similar to TRAPPIST-1c if all stars analysed in this study are orbited by a planet of the tested type. Therefore, it is not surprising that only one system was detected.

Using the pipeline, we would have found the planets b and c automatically, which validates the pipeline. An indication for the existence of planet c would have been noticed first because it transits during the second observation night and its transit is clearly visible and easily identified by the pipeline.

In accordance with previous results \citep[e.g.][]{dressing_2015, Demory_2016}, we find that hot mini-Neptunes are likely to be rare around cool stars since the probability of missing these planets is low (cf. Fig. \ref{fig:expmap2}) and we found none in the data set. Figs. \ref{fig:expmap1} and \ref{fig:expmap2} further indicate that we are sensitive to Earth-sized planets with short periods of the order of up to three days only.

\citet{He_2017} performed a planet injection analysis on brown dwarf light curves acquired by the \textit{Spitzer Space Telescope}. Their study lead to an expectation value for the number of discovered planets of (0.6, 0.81, 0.83, 0.85, 0.89) within the radius bins (1$\pm$0.25, 1.5$\pm$0.25, 2$\pm$0.25, 2.5$\pm$0.25, 3$\pm$0.25) $R_{\oplus}$  for an orbital period between 1.28 and 3 days. We expect to find (0.24, 0.48, 0.59, 0.62, 0.64) planets in the same parameter range. They observed 44 brown dwarfs with a median observation time of 20.9 hours. Thus, they have observed their targets for a shorter amount of time. However, they count one well-observed planetary transit as a detection while we require the detection of at least two transits. Furthermore, their observations are almost continuous while ours can be quite spread out in time, which can adversely affect the detection probability. The expectation value in \citet{He_2017} flattens out for radii above 1.5~$R_{\oplus}$ while this is the case for radii above 2 $R_{\oplus}$ in our analysis. This is likely the case because the \textit{Spitzer} light curves do not contain short-term variations due to, for example, precipitable water vapour.

Injection-recovery tests by \citet{Demory_2016} on 189 late M dwarfs observed in \textit{K2}'s Campaigns 1--6 yielded a recovery efficiency of 10 per cent for planets similar to TRAPPIST-1b. \textit{K2} has been monitoring each campaign field quasi-continuously for approximately 80 days, which is much longer than our typical summed exposure time of 50 hours. Nevertheless, we got a detection probability of 30 per cent in the period and radius bin of TRAPPIST-1b. Possible reasons for this difference are the fact that late M dwarfs are faint in the \textit{Kepler} passband and that the \textit{Kepler} cadence of 30 minutes can smear out the transits. Furthermore, \citet{Demory_2016} used a criterion based on the BLS period, which is likely to yield a lower detection probability for active stars compared to our detection criterion defined in Section \ref{identificationcriterion}.

\subsection{Multiplanetary systems}
Planetary systems like TRAPPIST-1, consisting of multiple, similarly sized planets in edge-on, coplanar orbits, produce comparably deep transits with the characteristic, sharp transit shape. A human analysing a light curve of such a planetary system would connect the transits produced by different planets and further observe this target star. The same is true for the BLS algorithm, as in the case of TRAPPIST-1, it first connects the transits of TRAPPIST-1c and TRAPPIST-1g before finding the transits of TRAPPIST-1b, which are very clear compared to the first two transits. Thus, multiplanetary systems similar to TRAPPIST-1 are ideal targets for surveys such as TRAPPIST-UCDTS. Generally, multiplanetary coplanar systems lead to a higher identification probability.
If all surveyed stars are orbited by two planets with planetary radii between 1 and 1.3 $R_{\oplus}$ at randomly drawn inclinations, one of them with a period between 1 and 1.4 days and the other with a period between 1.8 and 2.3 days, we estimate the probability of detecting at least one of the planets to 69 per cent (cf. Fig. \ref{fig:expmap2}). If coplanarity is assumed, however, the innermost planet is the dominant contributor to the detection probability which is thus above 57 per cent for the mentioned case.
We must conclude in both cases that a close-in planetary system, consisting of one or more planets, is relatively likely to be discovered in a survey like TRAPPIST-UCDTS, if these systems are common around UCDs.
Therefore, we expect to find some multi-planet systems in surveys such as SPECULOOS \citep{burdanov_2018,delrez_speculoos_2018, gillon2018, jehin_2018}, consisting of multiple telescopes fully dedicated to the search of exoplanets orbiting UCDs and achieving a higher photometric precision than TRAPPIST-South, if compact coplanar planetary systems are frequent. The latter is supported by evidence that compact multiple systems are relatively common around mid-type M dwarfs. Considering planets with periods of less than 10 days, \citet{Muirhead_2015} found that $21^{+7}_{-5}$ per cent of all mid-type M dwarfs host compact multiple systems, while \citet{Hardegree_2019} estimated an occurrence rate of $44^{+45}_{-33}$ per cent. We accompany these results with an estimate of the lower limit on the occurrence rate between 7 and 14 per cent for planets with a radius in [1,1.3] $R_\oplus$ and the period in [1.4,1.8] days hosted by UCDs.

\subsection{Transit detection challenges}

With few exceptions \citep{dalal_2019,huber_2013}, measurements of the obliquities of stars hosting multiple planets \citep[e.g.][]{hirano_2020,sanchis_2012,hirano_2012,albrecht_2013,chaplin_2013} suggest that  the axis of rotation of the host star is likely to be close to parallel to the orbital axis of its planets. Stars with close-in transiting planets, are therefore unlikely to exhibit a quiet light curve since the starspot and the inhomogeneous cloud coverage expected on the stellar surface of late-type M dwarfs and brown dwarfs \citep{metchev_2015,goldman_2005} will constantly change as seen from Earth due to stellar rotation. This is critical given the fast rotation of many brown dwarfs and very-low-mass stars of the order of one day \citep[e.g.][]{Irwin_2011,scholz_2016}. Additionally, precipitable water vapour can cause variability in the light curves \citep{Bailer-Jones_2003, Murray_2020} further complicating the confirmation of a transit candidate.

Therefore, it is likely that some transits, especially grazing transits, have not been identified by eye if the time scale and amplitude of the brightness variations during the respective night are comparable to those of planetary transits. This indicates that a rigorous computational search for periodic transit signals with simultaneous variability correction is necessary.

\section*{Acknowledgements}
We thank the anonymous referee for their thorough review which helped elevate the quality of this work. FL gratefully acknowledges a scholarship from the Fondation Zd\u{e}nek et Michaela Bakala. The research leading to these results has received funding from the European Research Council under the FP/2007-2013 ERC Grant Agreement n$^{\circ}$ 336480 and from the ARC grant for Concerted Research Actions, financed by the Wallonia-Brussels Federation. TRAPPIST is funded by the Belgian Fund for Scientific Research (Fonds National de la Recherche Scientifique, FNRS) under the grant FRFC 2.5.594.09.F, with the participation of the Swiss National Science Foundation (SNF). MG and EJ are F.R.S.-FNRS Senior Research Associates. AM acknowledges support from the senior Kavli Institute Fellowships. This research has made use of the SIMBAD database, operated at CDS, Strasbourg, France.

\section*{Data availability}
Data will be available via VizieR at CDS.


\bibliographystyle{mnras}
\bibliography{export-bibtex2} 




\appendix

\section{Occurrence rate}\label{sec:occurrence}

In the following, we summarise the stellar and planetary parameters ($R, P, M_{\ast_j}, R_{\ast_j}$) of the potential star-planet system $j$ in $\Theta_j$. These parameters for the whole survey are thus encoded in $\Vec{\Theta}$. Additionally, we abbreviate TRAPPIST-1 as T1. $\Theta_{T1b}$ thus stands for the planetary radius and period of TRAPPIST-1b and the stellar radius and mass of its host star TRAPPIST-1.  

The probability of detecting a planet around star $j$, if it hosts a planet, can be derived using
\begin{equation}
    \mathbb{P}(\text{detect planet} \: | \: \Theta_j, N_j = 1) = \mathbb{P}_{\text{identify}_j} \cdot \mathbb{P}_{\text{geometry}_j} = f(\Theta_j).
\end{equation}
 We use $\Vec{N}$ to describe the potentially hidden planetary population of the target stars. $N_j$ is set to 1 if there is a planet around star $j$ and it is set to 0 otherwise.
We assume that the false positive probability, i.e. the probability of detecting a planet around star $j$ if it does not host a planet, is equal to 0:
\begin{equation}
    \mathbb{P}(\text{detect planet} \: | \: \Theta_j, N_j = 0) = 0.
\end{equation}
Additionally, we assume that we correctly infer the existence of the planet if the detection criterion in Section \ref{identificationcriterion} is met. 
We set the probability that there is a planet orbiting planet $j$ to:
\begin{equation}
    \mathbb{P}(N_j = 1) = r.
\end{equation}
Analogously, 
\begin{equation}
    \mathbb{P}(N_j = 0) = 1-r.
\end{equation}

Next, we have to marginalise over detection probabilities since we do not know the planetary parameters of the potential unknown planets. For this, we have to make an assumption about the occurrence rates of planets within the chosen parameter range. In case 1) we marginalise using a uniform prior. In this case, we assume that we have no indication about the occurrence rate of planets orbiting UCDs. The parameter space covering short orbits and large radii, where we have a high detection probability but planets seem to be rare, is thus treated the same as short orbit Earth-sized planets which might be more common.
The occurrence rates for UCDs are unknown, but there are occurrence rate estimates for M dwarfs in general. In case 2), we use the occurrence rates from M dwarfs derived from the \textit{Kepler} mission data by \cite{dressing_2015}, normalised over the parameter space, as our new prior. We thus essentially exclude hot Neptunes. In both cases, we set the minimal planetary radius $R_{\text{min}}$ to 1~$R_{\oplus}$, the maximal radius $R_{\text{max}}$ to 4~$R_{\oplus}$, the shortest orbital period $P_{\text{min}}$ to 0.9~days, and the longest period $P_{\text{max}}$ to 3 days to match the bin choice in \cite{dressing_2015} figure 12.\\
Case 1) Uniform prior:
To marginalise over the unknown planetary radius and period, we integrate over a uniform prior $P(R,P)$:
\begin{equation}
    F_j = \int_{R_{\text{min}}}^{R_{\text{max}}} \int_{P_{\text{min}}}^{P_{\text{max}}} \frac{f(\Theta_j) dR dP}{\left(R_{\text{max}}-R_{\text{min}} \right) \left(P_{\text{max}} - P_{\text{min}} \right)}.
\end{equation}
\\
Case 2) Normalised \textit{Kepler} occurrence rates for M dwarfs as prior:
To marginalise over the unknown planetary radius and period we integrate over a uniform prior $P(R,P)$:
\begin{equation}
    F_j = \int_{R_{\text{min}}}^{R_{\text{max}}} \int_{P_{\text{min}}}^{P_{\text{max}}} f(\Theta_j) p(\Theta_j) dR dP.
\end{equation}
where $p(\Theta_j)$ are the occurrence rates as in \cite{dressing_2015} figure~12 but normalised over our integration range since we want to use the parameter $r$ as the number of planets per star within this range. In both cases we assume our stellar masses and radii to be correct.

For both cases we continue as follows. Our data $D$ is the survey outcome that we initially find T1b but no other planets. The assumption here is, that T1b is the true trigger for the detection of the other TRAPPIST-1 planets.
\begin{itemize}
    \item A: detect T1b in the T1 light curve.
    \item B: detect no other planets in the other light curves.
\end{itemize}
\begin{equation}
    \mathbb{P}(\text{A} \: | \: r) \: = \: r f(\Theta_{T1b})[N_{T1b} = 1] 
\end{equation}
  
\begin{equation}
    \mathbb{P}(\text{B} \: | \: r, \Vec{N}) \: = \: \prod_{j \: \in \: \text{S}}\left(\left(1-r F_j\right)[N_j=1] + [N_j=0] \right)
\end{equation}

where S is the set of all target stars except T1, and
\begin{equation}
    [\text{x}] = \begin{cases}
1 &\text{if x is true}\\
0 &\text{otherwise.}
\end{cases}
\end{equation}

For a given planetary population around the target stars, the probability of detecting exactly T1b and no other planets is then equal to
\begin{equation}
    \mathbb{P}(D \: | \: r, \Vec{N}) = \mathbb{P}(\text{A} \: | \: r, \Vec{N}) \cdot \mathbb{P}(\text{B} \: | \: r, \Vec{N}).
\end{equation}

Since the true planetary population around the stars (except in the case of T1), is unknown, we have to sum over all possible configurations and assign the respective probability to each configuration. We assume again a fixed occurrence rate $r$ for all stars and for all considered planetary and stellar parameters.
The probability of our data is thus:
\begin{equation}
    \mathbb{P}(D \: | \: r) = \sum_{\Vec{N}} \mathbb{P}(D \: | \: r, \Vec{N}) \cdot \mathbb{P}(\Vec{N} \: | \: r)
\end{equation}
with
\begin{equation}
    \mathbb{P}(\Vec{N} \: | \: r) = \prod_j r^{[N_j=1]} \cdot (1-r)^{[N_j=0]}.
\end{equation}
To summarise, $\mathbb{P}(D \: | \: \Vec{\Theta}, r)$ is equal to
\begin{equation}
    \mathbb{P}(D \: | \: r) = r f(\Theta_{T1b})[N_{T1b} = 1] \cdot \sum_{\Vec{N}} \mathbb{P}(B \: | \: r, \Vec{N}) \cdot \mathbb{P}(\Vec{N} \: | \: r),
\end{equation}
which can be written out as:
\begin{multline}
\mathbb{P}(D \: | \: r) = \\
r f(\Theta_{T1b})[N_{T1b} = 1] \cdot \\
\{ (1-r)^{n_{\text{stars}}} +\\
\sum_{j \: \in \: \text{S}} (1-F_j) \cdot r (1-r)^{n_{\text{stars}}-1} +\\
\sum_{j \: \in \: \text{S}}\left( (1-F_j) \cdot \sum_{k \: \in \: \text{S} \setminus j} (1-F_k) \cdot r^2 (1-r)^{n_{\text{stars}}-2}\right) +\\
...\}.\\
\end{multline}
This can be simplified to:
\begin{equation}
    \mathbb{P}(D \: | \: r) = 
    r f(\Theta_{T1b}) \cdot \prod_{j \: \in \: \text{S}} 
    \left( \left(1-r\right) + r\left(1-F_j \right) \right),
\end{equation}
which is equal to

\begin{equation}
    \mathbb{P}(D \: | \: r) = 
    r f(\Theta_{T1b}) \cdot \prod_{j \: \in \: \text{S}} 
    \left( 1-r F_j \right) .
\end{equation}


\bsp	
\label{lastpage}
\end{document}